\documentclass{article}
\usepackage[german, english]{babel}
\usepackage[a4paper, left=2.5cm, right=2.5cm, top=3.5cm, bottom=3cm]{geometry}
\usepackage{amssymb}
\usepackage{amsmath}
\usepackage{bbm}
\usepackage{amscd}
\usepackage{latexsym}
\usepackage{graphicx}
\usepackage{ifthen}
\usepackage{epsfig}
\usepackage{setspace}
\usepackage{subcaption}
\usepackage{hyperref}
\usepackage{dsfont}
\usepackage{mathrsfs}
\usepackage{color}
\usepackage{array}
\newcolumntype{P}[1]{>{\centering\arraybackslash}p{#1}}
\usepackage{makecell}
\usepackage[toc,page]{appendix}
\hypersetup{
 colorlinks=true,
 linkcolor=blue,
 citecolor=magenta,
}
\catcode`@=11
\renewcommand{\section}{\@startsection{section}{1}{0pt}{\medskipamount}
{\medskipamount}{\large\bf}}
\numberwithin{equation}{section}
\catcode`@=12

\newcommand{\N}{\mathds N}
\newcommand{\R}{\mathds R}

\newcommand{\Acal}{{\cal A}}
\newcommand{\Fcal}{{\cal F}}
\newcommand{\Ecal}{{\cal E}}
\newcommand{\Bcal}{{\cal B}}

\newcommand{\tY}{{\widetilde{Y}}}

\def\im{\mathrm{i}}
\def\ep{\mathrm{e}}
\def\pa{\mbox{$\partial$}}
\def\diff{\mathrm{d}}

\def\sfrac#1#2{{\textstyle\frac{#1}{#2}}}
\def\]{\right]}
\def\[{\left[}
\def\){\right)}
\def\({\left(}
\def\>{\rangle}
\def\<{\langle}
\def\+{\dagger}
\def\we{{\wedge}}
\def\={\ =\ }

\def\und{\quad\textrm{and}\quad}
\def\with{\quad\textrm{with}\quad}

\def\3j#1#2#3#4#5#6{\begin{pmatrix} #1&#2&#3\\#4&#5&#6 \end{pmatrix}}
\def\s3j#1#2{\begin{pmatrix} #1\\#2 \end{pmatrix}}

\DeclareMathOperator{\sech}{sech}

\begin{document}

\title{\bf\huge Conserved charges for \\ rational electromagnetic knots}
\date{~}

\author{
{\Large Lukas Hantzko}, \ 
{\Large Kaushlendra Kumar}\footnote{Corresponding author: kaushlendra.kumar@itp.uni-hannover.de} \ and \
{\Large Gabriel Pican\c{c}o Costa}
\\[24pt]
Institut f\"ur Theoretische Physik \\
Leibniz Universit\"at Hannover \\ 
Appelstra{\ss}e 2, 30167 Hannover, Germany
\\[24pt]
} 

\clearpage
\maketitle
\thispagestyle{empty}

\begin{abstract}
\noindent\large
We revisit a newfound construction of rational electromagnetic knots based on the conformal correspondence between Minkowski space and a finite $S^3$-cylinder. We present here a more direct approach for this conformal correspondence based on Carter--Penrose transformation that avoids a detour to de Sitter space. The Maxwell equations can be analytically solved on the cylinder in terms of $S^3$ harmonics $Y_{j;m,n}$, which can then be transformed into Minkowski coordinates using the conformal map. The resultant ``knot basis" electromagnetic field configurations have non-trivial topology in that their field lines form closed knots. We consider finite, complex linear combinations of these knot-basis solutions for a fixed spin $j$ and compute all the 15 conserved Noether charges associated with the conformal group. We find that the scalar charges either vanish or are proportional to the energy. For the non-vanishing vector charges, we find a nice geometric structure that facilitates computation of their spherical components as well. We present analytic results for all charges for up to $j{=}1$. We demonstrate possible applications of our findings through some known previous results. 
\end{abstract}

\newpage
\setcounter{page}{1} 

\section{Introduction and summary}
\noindent
Topologically non-trivial solutions to source-free Maxwell's equations on Minkowski space $\R^{1,3}$ were first discovered by Ra\~{n}ada \cite{Ranada89} based on the Hopf fibration. Other methods for constructing such linked and knotted rational solutions involving Bateman's complex Euler potentials, conformal inversion and Penrose twistors have been found since then (see \cite{ABT17} for a review). Key features characterizing these solutions are helicity (conserved under conformal transformations) and Noether charges associated with the conformal group $SO(2,4)$ of $\R^{1,3}$ \cite{IB08,HSS15}. Quite interestingly, some simple EM knot configurations, including the one with figure-8 topology, have been produced in the laboratory using laser beams with knotted polarization singularities \cite{LSMetal18}. With such a discovery, this line of research no longer remains just a theoretical endeavor; more complicated knot configurations discussed here, or elsewhere, could be realized experimentally in the near future. Trajectories of charged particles in the background of these knotted electromagnetic fields have also been studied recently \cite{KLP22}.

It was shown in \cite{LZ18} that a complete family of such rational knot solutions on $\R^{1,3}$, labelled by the so called left-right $S^3$ harmonics $Y_{j;m,n}$, can be constructed; any finite-action rational Maxwell solution can be expanded in terms of these basis solutions. The construction relies on the conformal correspondence between the de Sitter space dS$_4$ and $\R^{1,3}$ via a finite Lorentzian $S^3$-cylinder over conformal time $\tau$. Here, we establish this conformal correspondence between the cylinder and $\R^{1,3}$ more directly using Carter--Penrose transformation. These ``knot basis" solutions are first obtained on the cylinder by employing the $SO(4)$ isometry of $S^3$ coupled with the right-multiplication action of $SU(2)$ on $S^3$ (being the group manifold of $SU(2)$) and are later transferred to $\R^{1,3}$ via the conformal map. In another paper \cite{KL20}, co-authored by one of us, general linear combination of such ``knot basis" solutions in terms of arbitrary complex coefficients $\Lambda_{j,m,n}$ was considered and, for a fixed $j$, its helicity was found to be proportional to the conserved energy. 

Here, we reconsider these generic solutions for a fixed $j$ and proceed to compute the remaining charges associated with the conformal group, namely momenta, angular momenta, boosts, dilatation and special conformal transformation (SCT) charges. The charge densities are evaluated on de Sitter space at $\tau=0$ (without loss of generality, since they are conserved) where considerable simplifications occur demonstrating the usefulness of this ``de Sitter method". It is found that the dilatation charge vanishes while the scalar SCT charge $V_0$ is proportional to the energy $E$. Furthermore, the boost charges $K_i$ vanish and the vector SCT charges $V_i$ are proportional to the momentum charges $P_i$. Interestingly, for the vector charge densities viz. momenta $p_i$, angular momenta $l_i$ and vector SCT $v_i$ we find that a one-form constructed with them on the spatial slice $\R^3 \hookrightarrow \R^{1,3}$, e.g. $p_i\diff x^i$ is proportional to a similar one on de Sitter space. This correspondence allows us to compute additional charges $(p_r,p_\theta,p_\phi)$ by the action of such one-forms on spherical vector fields $(\pa_r,\pa_\theta,\pa_\phi)$. At $j=0$ it turns out that there are only four independent nonzero charges: the energy $E$ and the momenta $P_i$. The situation for higher spin $j$ is more complicated, but some of the components of the charges in spherical coordinates are found to vanish for arbitrary $j$. The action of the $\frak{so(3)}$ generators $\mathcal{D}_a$ on the indices of $\Lambda_{j,m,n}$ can easily be obtained for a fixed $j$ owing to the $SO(4)$ isometry. This allows for an action of these generators on the charges. For (the Cartesian components of) the vector charges this action is found to inherit the original $\frak{so(3)}$ Lie algebraic structure, as expected. We also compute the correct coefficients $\Lambda_{0;0,n}$ corresponding to two very interesting generalizations of the Hopfian solution obtained via Bateman's construction in \cite{HSS15}, which allow us to validate our generic formulae of these charges. Further works in this direction could be to relate these new knot solutions and their conserved charges to other known field configurations in \cite{HSS15}, such as plane waves or to study Fourier analysis of general electro-magnetic knot configurations to understand single mode behaviour. 

The paper is organized as follows. In Section \ref{sec2}, we first establish the conformal map between Minkowski space and the Lorentzian $S^3$-cylinder using Carter--Penrose transformation and review the necessary computational tools related to $S^3$, including its harmonics. We then construct the rational Maxwell solutions on the cylinder in terms of the $S^3$ harmonics and illustrate some knotted field configurations using figures. Section \ref{sec3}, at first, reviews the Noether current and associated charges arising from the action of the conformal group on free Maxwell theory and presents a computational strategy. The $15$ conformal charges are then computed in four subsections and the action of $\frak{so(3)}$ generators $\mathcal{D}_a$ on some of the vector charges is analyzed. Finally, in Section \ref{sec4} we validate the formulae of the charges computed in Section \ref{sec3} using other known results.

\section{Construction of rational electromagnetic knots}
\label{sec2}
\noindent
We present below the construction of rational electromagnetic knots by completely skipping the detour to de Sitter space in contrast with \cite{LZ18}, where these solutions were first presented. For a clear presentation, we split the construction in three parts.

\subsection{Minkowski to cylinder: Carter--Penrose transformation}
\noindent
The metric on the Minkowski space $\R^{1,3}$ in polar coordinates is given by
\begin{equation}\label{metricMink}
    \diff s_\text{Mink}^2 \= -\diff t^2 + \diff r^2 + r^2\,\diff\Omega_2^2\ ;\qquad \diff\Omega_2^2 \= \diff\theta^2 + \sin^2\theta\,\diff\phi^2 ,
\end{equation}
\begin{figure}[h!]
\centering
\includegraphics[width = 0.3\paperwidth]{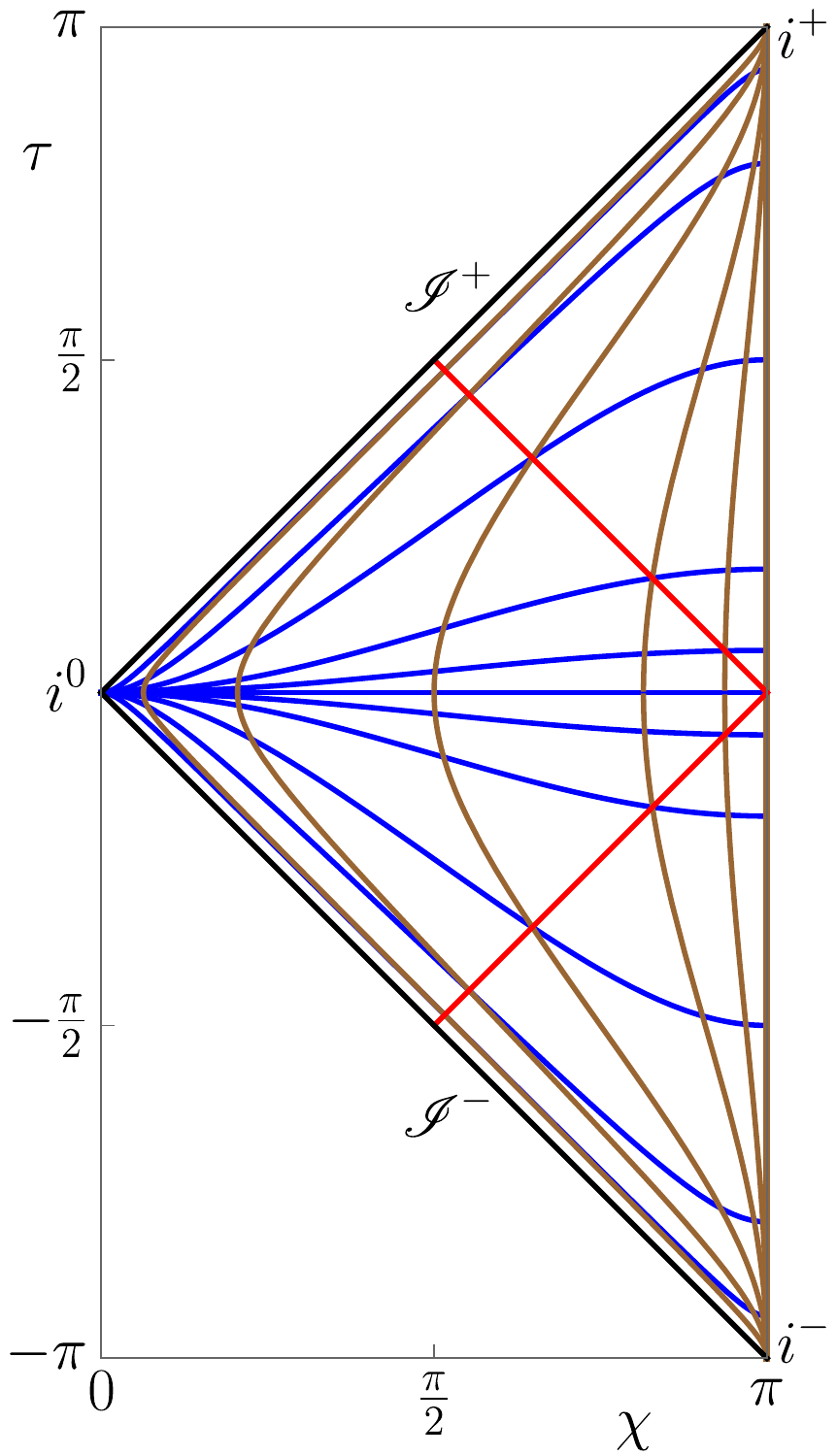}
\caption{
Penrose diagram of Minkowski space~$\R^{1,3}$. Each point hides a two-sphere $S^2\ni\{\theta,\phi\}$.
Blue curves indicate $t{=}\textrm{const}$ slices while brown curves depict the world hypersurface of $r{=}\textrm{const}$ spheres.
The lightcone of the Minkowski-space origin is drawn in red.
}
\label{fig1}
\end{figure}
where $-\infty < t < \infty,\ 0 \leq r < \infty,\ 0 \leq \theta \leq \pi$, and $0 \leq \phi \leq 2\pi$. We first employ the lightcone coordinates $(u,v)$ to transform the metric in the following way
\begin{equation}\label{transf1}
\begin{aligned}
    &u\ :=\ t-r\ ,\quad v\ :=\ t+r\ ;\quad  -\infty < u \leq v < \infty\  \\[2pt]
    &\implies \diff s_\text{Mink}^2 \= -\diff v\, \diff u + \sfrac14 (v-u)^2\,\diff\Omega_2^2\ .
\end{aligned}
\end{equation}
In the second step, we compactify the spacetime with the help of the coordinate (U,V) as follows:
\begin{equation}\label{transf2}
\begin{aligned}
   &U\ :=\ \arctan u\ ,\quad V\ :=\ \arctan v\ ;\quad  -\sfrac\pi2 < U \leq V < \sfrac\pi2\  \\[2pt]
    &\implies \diff s_\text{Mink}^2 \= \frac{1}{4\cos^2V\cos^2U}\left[-4\,\diff V\, \diff U +  \sin^2(V-U)\,\diff\Omega_2^2\right]\ .
\end{aligned}
\end{equation}
Finally, we rotate the coordinate system back using $(\tau,\chi)$ to obtain the desired form of the metric:
\begin{equation}\label{transf3}
\begin{aligned}
   &\tau\ :=\ V + U\ ,\quad \chi\ :=\ \pi + U - V\ ;\quad  0 < \chi \leq \pi\ , |\tau| < \chi\  \\[2pt]
    &\implies \diff s_\text{Mink}^2 \= \gamma^{-2}\left[-\diff \tau^2 + \diff\chi^2 +  \sin^2\chi\,\diff\Omega_2^2\right]\ ; \quad \gamma \= \cos\tau - \cos\chi\ .
\end{aligned}
\end{equation}
We realize that the Minkowski metric is conformally equivalent to the metric on a Lorentzian cylinder $\mathcal{I}\times S^3\ ;\ \mathcal{I}=(-\pi,\pi)$ with a conformal factor that can be recasted in terms of Minkowski coordinates using the above transformations (\ref{transf1}-\ref{transf3}):
\begin{equation}\label{metricCyl}
      \diff s_\text{cyl}^2\ :=\ -\diff\tau^2 + \diff\Omega_3^2 \= \gamma^2\, \diff s_\text{Mink}^2\ ;\quad \gamma = \frac{2\ell^2}{\sqrt{4\,t^2\,\ell^2 + (r^2-t^2+\ell^2)^2}}\ ,
\end{equation}
where we have made $\gamma$ dimensionless using a global scale factor $\ell$, chosen to be the de Sitter radius so as to make the connection with previous works apparent. We depict in Figure \ref{fig1} the causal structure of the Minkowski spacetime (preserved under a conformal transformation) using the Penrose diagram. Note that the spatial infinity $i^0$ (where $t=0$ and $r\rightarrow\infty$) and the future/past temporal infinities $i^\pm$ (where $r=0$ and $t\rightarrow\pm\infty$) are not included in the figure. Furthermore, the boundaries of the Penrose diagram are (a) $r=0$ line, (b) past null infinity $\mathscr{I}^-$ corresponding to $r-t\rightarrow\infty$, $r+t =$ constant and (c) future null infinity $\mathscr{I}^+$ corresponding to $r+t\rightarrow\infty$, $r-t =$ constant.
The natural embedding coordinates $\omega_{_A}$ with $A=1,2,3,4$ of $S^3\hookrightarrow\R^4$ satisfying $\omega_{_A}^2=1$ has a natural parametrization in terms of hyperspherical coordinates $(\chi,\theta,\phi)$ of \eqref{transf3}:
\begin{equation}\label{omegas}
   \omega_a \= \sin\chi\;\hat{\omega}_a\ ,\ \omega_4 \= \cos\chi\ , \with \left(\hat{\omega}_1,\, \hat{\omega}_2,\,\hat{\omega}_3 \right) \= \left( \sin\theta\cos\phi,\, \sin\theta\sin\phi,\, \cos\theta \right)\ .
\end{equation}
It is not difficult to obtain the following useful relations
\begin{equation} \label{tr2TauChi}
   \gamma\,t \= \ell\,\sin\tau \quad\und\quad
   \gamma\,r \= \ell\,\sin\chi\ ,
\end{equation}
which can be used to find the Jacobian of transformation between the coordinates $y^m\in\{\tau,\chi,\theta,\phi\}$ and $x^{\mu}\in\{t,r,\theta,\phi\}$:
\begin{equation} \label{Jacobian}
   \bigl(J^m_{\ \ \mu}\bigr) \ :=\ \frac{\pa(\tau,\chi,\theta,\phi)}{\pa(t,r,\theta,\phi)}
   \= \frac{1}{\ell}\biggl(\begin{matrix} p & -q \\[4pt] q & -p \end{matrix} \biggr) \oplus\mathds{1}_2\ 
   \with \biggl\{ \begin{array}{l}
   p\= \sfrac{\gamma^2}{2\ell^2}\,(r^2{+}t^2{+}\ell^2) \= 1{-}\cos\tau\cos\chi \\[4pt]
   q\= \sfrac{\gamma^2}{\ell^2}\,t\,r \= \sin\tau\sin\chi \end{array}\biggr\}\ .
\end{equation}

\subsection{Structure of three-sphere and its harmonics}
\noindent
The presence of $S^3$ in the metric \eqref{metricCyl} is an added advantage, which can be exploited to write a $SO(4)$-invariant vector potential and obtain the corresponding electromagnetic fields by solving Maxwell's equations on the Lorentzian cylinder. These quantities can be later exported to the Minkowski spacetime owing to the conformal invariance of the vacuum Maxwell equations in four dimensions. To this end, we start with the group $SO(4)$, which is isomorphic to two copies of $SU(2)$ (up to a $\mathds{Z}_2$ grading). Each of these $SU(2)$ has a group action that generates a left (right) multiplication (a.k.a. translation) on $S^3$. This can easily be checked using the map
\begin{equation}\label{map}
   g:\; S^3 \rightarrow SU(2);~~ (\omega_1,\ \omega_2,\ \omega_3,\ \omega_4)\, \mapsto\, -i
     \begin{pmatrix}
     \beta & \alpha^* \\ \alpha & -\beta^*
     \end{pmatrix} 
   ~\textrm{with}~\ \alpha := \omega_1 + i\omega_2\ ~\textrm{and}~\ \beta := \omega_3 +i\omega_4\ .
\end{equation} 
This parametrization of $g$ ensures that the identity element $e = \mathds{1}_2$ of the group $SU(2)$ can be obtained from $(0,0,0,1)$, i.e. the North pole of $S^3$. It is well known that $S^3$ is the group manifold of $SU(2)$. Keeping this in mind, we consider the Cartan one-form
\begin{equation} \label{Cartan1}
   \Omega_l(g)\, :=\, g^{-1}\,\diff g \= e^a\,T_a,~~\textrm{with}~~ T_a = -i\sigma_a
\end{equation} 
being the $SU(2)$ generators. The left-invariant one-form\footnote{Named so because they remain invariant under the dragging induced by the left SU(2) multiplication.} $e^a$ can, alternatively, be expressed using the so-called self-dual 't Hooft symbol $\eta^a_{_{\ BC}}$:
\begin{equation}\label{one-form}
   e^a \= -\eta^a_{_{\ BC}}\, \omega_{_B}\, \diff \omega_{_C} \with \eta^a_{\ bc} \= \varepsilon_{abc} \und \eta^a_{\ b4} \= -\eta^a_{\ 4b} \= \delta^a_b\ .
\end{equation}
They satisfy the following useful identities
\begin{equation}\label{MaurerCartan}
\delta_{ab}\, e^a\, e^b \= \diff\Omega_3^2 \quad\und\quad \diff e^a + \varepsilon_{abc}\, e^b\wedge e^c \= 0.
\end{equation} 
The left-invariant vector fields $L_a$ generating the right translations are dual to $e^a$ and are given by
\begin{equation}\label{leftVF}
   L_a \= -\eta^a_{_{\ BC}}\,\omega_{_B}\,\sfrac{\partial}{\partial \omega_{_C}} \qquad\Rightarrow\qquad \left[ L_a , L_b \right] \= 2\,\varepsilon_{abc}\,L_c\ .
\end{equation}
In a similar way, the right-invariant vector fields $R_a$ generating the left translations are given by
\begin{equation}\label{rightVF}
   R_a \= -\tilde{\eta}^a_{_{\ BC}}\,\omega_{_B}\,\sfrac{\partial}{\partial\omega_{_C}} \qquad\Rightarrow\qquad
   \left[ R_a , R_b \right] = 2\,\varepsilon_{abc}\,R_c\ ,
\end{equation}
where the anti-self-dual 't Hooft symbols $\tilde{\eta}^a_{_{\ BC}}$ are obtained from \eqref{one-form} by flipping the $B,C\!=4$ sign. The stability subgroup $SO(3)\subset SO(4)$ under the group action on the coset space $SO(4)/SO(3) \simeq S^3$ is generated by
\begin{equation}\label{so3rot}
    \mathcal{D}_a\ :=\ L_a + R_a = -2\, \varepsilon_{abc}\,\omega_b\, \sfrac{\partial}{\partial\omega_c} \quad\implies\quad [\mathcal{D}_a,\mathcal{D}_b] = 2\, \varepsilon_{abc}\, \mathcal{D}_c\ .
\end{equation}
Furthermore, the vector fields $L_a$ and $R_a$ act on the one-forms $e^a$ via their Lie derivative\footnote{One can use, e.g., the Cartan formula: $\mathcal{L}_{X} = \diff\circ\iota_X + \iota_X\circ\diff$ where $\mathcal{L}_{X}$ is the Lie derivative w.r.t $X$, $\diff$ is the exterior derivative and $\iota_X$ is the interior product.} yielding
\begin{equation}\label{LieAction}
    L_a\, e^b \= 2\,\varepsilon_{abc}\,e^c \quad\und\quad R_a\, e^b \= 0\ .
\end{equation}
We can now write the differential $\diff$ of the functions $f\in C^{\infty}\left(\mathcal{I}\times S^3\right)$ using $L_a$ as
\begin{equation}
   \diff f \= \diff\tau\; \partial_{\tau} f\ +\ e^a\,L_a f\ .
\end{equation}
Furthermore, the restriction of such functions on $S^3$ has a natural expansion in terms of left-right harmonics $Y_{j;m,n}$ with $2j\in\N_0$ and $m,n\in\{-j,-j+1,...,j\}$ that are eigenfunctions of Casimirs\footnote{This is same as the scalar Laplacian on $S^3$: $L^2 = R^2 = \bigtriangleup_3$.} $L^2 := L_aL_a,\ R^2 := R_aR_a$, and operators $L_3$, $R_3$:
\begin{align} \label{Yalgebra}
 \begin{split}
   -\sfrac14 L^2\,Y_{j;m,n} &\= -\sfrac14 R^2\,Y_{j;m,n} \= j(j{+}1)\,Y_{j;m,n}\ , \\
   \sfrac{\im}{2} R_3\, Y_{j;m,n} &\= m\, Y_{j;m,n}\ , ~ \sfrac{\im}{2} L_3\, Y_{j;m,n} \= n\, Y_{j;m,n}\ .
 \end{split}
\end{align}
The action of the corresponding ladder operators 
\begin{equation}
   L_\pm \= (L_1\pm\im L_2)/\sqrt{2} \quad\und\quad R_\pm \= (R_1\pm\im R_2)/\sqrt{2}
\end{equation}
on the harmonics $Y_{j;m,n}$ is, therefore, given by
\begin{equation}\label{LadderAction}
   \sfrac{\im}{2}\,R_\pm\,Y_{j;m,n} \= \sqrt{(j{\mp}m)(j{\pm}m{+}1)/2}\,Y_{j;m\pm1,n} \quad\!\und\!\quad
   \sfrac{\im}{2}\,L_\pm\,Y_{j;m,n} \= \sqrt{(j{\mp}n)(j{\pm}n{+}1)/2}\,Y_{j;m,n\pm1}\ .
\end{equation}
The normalized harmonics $Y_{j;m,n}$ can be expanded in terms of functions $\alpha,\beta$ and their complex conjugates as
\begin{equation}\label{S3Harmonics}
    Y_{j;m,n}(\omega) \= \sqrt{\frac{2j+1}{2\pi^2}}\, \sum_{k=0}^{2j}\, (-1)^{m+n+k} \frac{\sqrt{(j+m)!(j-m)!(j+n)!(j-n)!}}{(n+m+k)!(j-n-k)!(j-m-k)!k!}\, \alpha^{n+m+k}\, \bar{\alpha}^{k}\, \beta^{j-m-k}\, \bar{\beta}^{j-n-k}\ .
\end{equation}
They satisfy the orthonormality condition
\begin{equation}\label{orthonormality}
    \int \diff^3\Omega_3\, Y_{j;m,n}\, \bar{Y}_{j';m',n'} \= \delta_{j,j'}\, \delta_{m,m'}\, \delta_{n,n'} \with \diff^3\Omega_3 \= \sin^2\chi\sin\theta\, \diff\chi\, \diff\theta\, \diff\phi\ .
\end{equation}
There also exist the so-called adjoint harmonics $\tilde{Y}_{j,l,M}$ \cite{KL20} that are related with the left-right harmonics $Y_{j;m,n}$ through the Clebsch--Gordan coefficients $C_{m,n}^{l,M}$ as
\begin{equation}
Y_{j;m,n} \= \sum\limits_{l=0}^{2j}\sum\limits_{M=-l}^{l}\,C_{m,n}^{l,M}\,\tY_{j;l,M}\ .
\end{equation}
These adjoint harmonics can be written in terms of the standard $S^2$ spherical harmonics $Y_{l,M}$ and the associated Legendre polynomials of the first kind $P_a^b$:
\begin{equation}\label{adjointY}
\tY_{j;l,M}(\chi,\theta,\phi) \= R_{j,l}(\chi)\,Y_{l,M}(\theta,\phi) \with
R_{j,l}(\chi) \= \im^{2j+l}\,\sqrt{\sfrac{2j+1}{\sin\chi}\sfrac{(2j+l+1)!}{(2j-l)!}}\, P_{2j+\frac12}^{-l-\frac12}(\cos\chi)\ .
\end{equation}
We can write down the cylinder one-forms in Minkowski coordinates using relations (\ref{omegas}-\ref{Jacobian}) in \eqref{one-form}
\begin{equation}\label{1formMap}
 \begin{aligned}
   \diff\tau\, =:\, e^\tau  &\= \sfrac{\gamma^2}{\ell^3}\bigl(
   \sfrac12(t^2{+}r^2{+}\ell^2)\,\diff t - t\,r\,\diff r \bigr) \\[2pt]
   e^a &\= \sfrac{\gamma^2}{\ell^3}\bigl(
   \hat{\omega}^a \bigl[ r\,t\,\diff t -\sfrac12(t^2{+}r^2{+}\ell^2)\,\diff r\bigr]
   - \sfrac12(t^2{-}r^2{+}\ell^2)\,r\,\diff\hat{\omega}^a -
   \ell\,r^2\varepsilon^{ajk}\hat{\omega}^j\diff\hat{\omega}^k \bigr)\ .
 \end{aligned}
\end{equation}
We further note down the expressions for the vector fields $L_a$ in terms of $S^3$ angles by using \eqref{omegas} in \eqref{leftVF} for later purposes:
\begin{equation}\label{Lfields}
 \begin{aligned}
   L_1 &\= \sin\theta\cos\phi\, \pa_\chi\ +\ (\cot\chi\cos\theta\cos\phi + \sin\phi)\,\pa_\theta\ -\ (\cot\chi\csc\theta\sin\phi - \cot\theta\cos\phi)\, \pa_\phi\ , \\
   L_2 &\= \sin\theta\sin\phi\, \pa_\chi\ +\ (\cot\chi\cos\theta\sin\phi - \cos\phi)\,\pa_\theta\ +\ (\cot\chi\csc\theta\cos\phi + \cot\theta\sin\phi)\, \pa_\phi\ , \\
   L_3 &\= \cos\theta\, \pa_\chi\ -\ \cot\chi\sin\theta\, \pa_\theta\ -\ \pa_\phi\ .
 \end{aligned}
\end{equation}

\subsection{Rational solutions to vacuum Maxwell equations}
\noindent
We start with the following ansatz \cite{KL20} for a real-valued Maxwell gauge potential $\mathcal{A}$ on the Lorentzian cylinder
\begin{equation} \label{Acalfull}
   \Acal \= \Acal_\tau(\tau,\omega)\,e^\tau\ +\ \sum_{a=1}^3 \Acal_a(\tau,\omega)\,e^a 
   \quad\with \omega\equiv\{\omega_{_A}\}\ ,
\end{equation}
which can be extended to the Minkowski space since the Maxwell's theory is conformally invariant here. We further impose the following ``temporal+Coulomb" gauge on this space
\begin{equation}\label{gaugeCond1}
   \Acal_\tau(\tau,\omega) = 0\quad\und\quad R_a\,\Acal_a (\tau,\omega) = 0\ .
\end{equation} 
Note that the first part in \eqref{gaugeCond1} is not the usual temporal gauge $A_t\!=0$. In fact, we can make use of the inverse Jacobian \eqref{Jacobian} while promoting the gauge potential to the Minkowski space $\Acal \!= \Acal_a\, e^a \!= A_\mu\,dx^\mu \!=A $ to get
\begin{equation}\label{gaugeCond2}
  0 = \Acal_\tau = \Acal \left(\partial_\tau \right) = A \left( \sfrac{\ell^2}{\gamma
  ^2}\left(p\,\partial_t + q\,\partial_r \right) \right) \implies \left(r^2+t^2+\ell^2\right)A_t + 2\,r\,t\,A_r = 0\ .
\end{equation}
The field strength $\cal{F}$ is easily obtained with this gauge potential using \eqref{MaurerCartan}:
\begin{equation}\label{2-form}
  \Fcal \= \diff\Acal \= 
  \pa_\tau\Acal_a\,e^\tau\we e^a + \bigl(\sfrac12 R_{[b}\Acal_{c]}-\Acal_a\,\varepsilon^a_{\ bc} \bigr)\,e^b\we e^c\ .
\end{equation}
The corresponding vacuum Maxwell equations $\diff*\Fcal=0$ read
\begin{equation} \label{Maxwell}
  \pa_\tau^2\Acal_a \= (R^2-4)\,\Acal_a + 2\,\varepsilon_{abc}R_b\Acal_c\ .
\end{equation}
Solutions to these coupled linear wave equations on $S^3$ were obtained in \cite{LZ18} as an infinite tower of basis solutions labelled by $j$ and divided into two categories: type-I with frequency $\Omega_j\!=2(j{+}1)$ and type-II with frequency $\Omega_j\!=2j$. We shall confine our discussions in this paper to type-I solutions. The corresponding results for type-II solutions can easily be obtained using a duality transformation (see \cite{LZ18}) where the basis solutions of type-I at level $j$ are related with the basis solutions of type-II at level $j{+}1$ via the exchange $m\leftrightarrow n$. We take a general complex linear combination of these basis solutions and obtain (see \cite{KL20}) the real-valued functions $\Acal_a(\tau,\omega)$ as 
\begin{equation}\label{AviaZ}
  \Acal_a(\tau,\omega) \= \sum_{2j=0}^\infty Z_{a}^j(\omega) \ \ep^{2(j+1)\im\tau} \ +\ \sum_{2j=0}^\infty \bar{Z}_{a}^j(\omega) \ \ep^{-2(j+1)\im\tau}\ ,
\end{equation}
where the complex angular functions $Z_a^j$ are expanded in terms of $(2j{+}1)(2j{+}3)$ arbitrary complex coefficients $\Lambda_{j;m,n}$ and basis solutions $Z_{*}^{j;m,n}(\omega) ~\textrm{with}~ *\in \{+,3,-\}$ as
\begin{equation}\label{basisChange}
    Z_{*}^j(\omega) \= \sum_{m=-j}^j \sum_{n=-j-1}^{j+1} \Lambda_{j;m,n}\ Z_{*}^{j;m,n}(\omega) \quad\with\quad Z_{\pm}^j\, :=\, \sfrac{1}{\sqrt{2}}\left( Z_1^j \pm\im Z_2^j \right)\ .
\end{equation}
The complex basis functions $Z_*^{j;m,n}$ are given in terms of harmonics $Y_{j;m,n}$ as
\begin{equation} \label{basisSoln}
 \left.\begin{aligned}
   Z_{+}^{j;m,n}(\omega) &\= \sqrt{(j{-}n)(j{-}n{+}1)/2} ~ Y_{j;m,n+1}(\omega) \\
   Z_{3}^{j;m,n}(\omega)\,&\= \sqrt{(j{+}1)^2-n^2} ~ Y_{j;m,n}(\omega) \\
   Z_{-}^{j;m,n}(\omega)   &\= -\sqrt{(j{+}n)(j{+}n{+}1)/2} ~ Y_{j;m,n-1}(\omega)
 \end{aligned}\right\}
  \with\biggl\{ \begin{array}{l}
   m\= -j,-j+1,...,j \\[4pt]
   n\= -j-1,-j,...,j+1 \end{array}\biggr\}\ .
\end{equation}
The expected harmonic expansion for $Z_a^j(\omega)$ is obtained by using  \eqref{basisChange} in \eqref{basisSoln} while inverting the basis from $(+,3,-)$ to $(1,2,3)$:
\begin{equation}\label{Zharmonics}
   Z_a^j(\omega) \= \sum_{m,n=-j}^j X_a^{j;m,n}\ Y_{j;m,n}(\omega) \quad\with\quad X_a^{j;m,n} \= \sum_{\tilde{n}=-j-1}^{j+1} C^{n,\tilde{n}}_{j,a}\, \Lambda_{j;m,\tilde{n}}\ ,
\end{equation}
where the real coefficients $C^{n,\tilde{n}}_{j,a}$ are determined from \eqref{basisSoln}.

The useful expression for the ``sphere-frame" electric and magnetic fields,
\begin{equation}\label{2FormEM}
    \Fcal \= \Ecal_a\,e^a \we e^\tau + \sfrac12\,\Bcal_a\,\varepsilon^a_{\ bc}\,e^b\we e^c\ 
\end{equation}
for a fixed spin~$j$ is obtained by eliminating $R_{[b}A_{c]}$ in \eqref{2-form} using \eqref{Maxwell} and employing
\begin{equation}
   \pa_\tau^2 \Acal^{(j)} = -\Omega_j^2\,\Acal^{(j)} \quad\und\quad
   R^2\,\Acal^{(j)} \= -4j(j{+}1)\,\Acal^{(j)}
\end{equation}
to get
\begin{equation}\label{curFields}
   \Ecal_a^{(j)} \= -\pa_\tau \Acal_a^{(j)} \quad\und\quad \Bcal_a^{(j)} \= -\Omega_j\,\Acal_a^{(j)}\ .
\end{equation}
\begin{figure}[h!]
\centering
   \includegraphics[width = 3cm, height = 5cm]{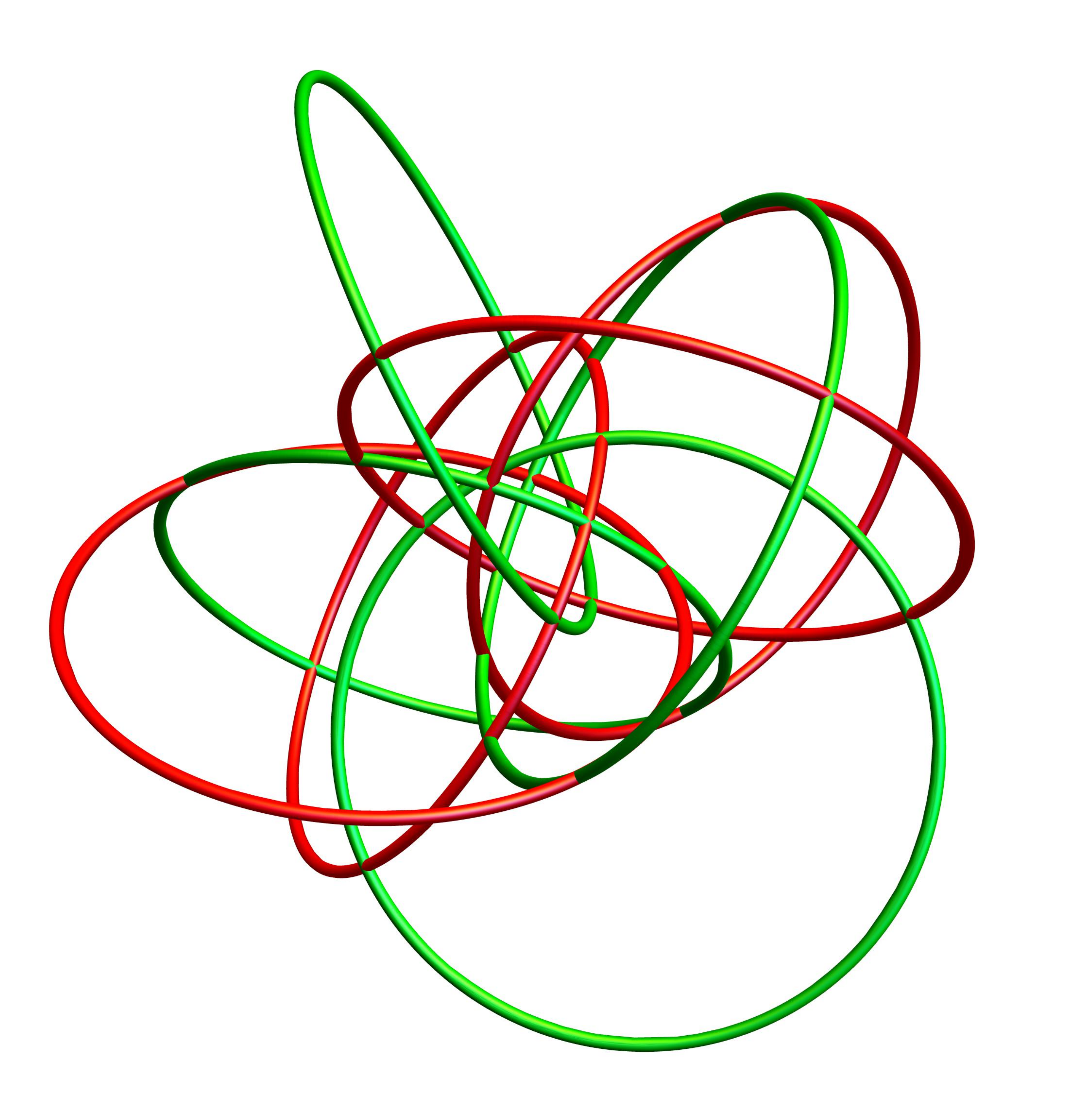}
   \includegraphics[width = 3cm, height = 5cm]{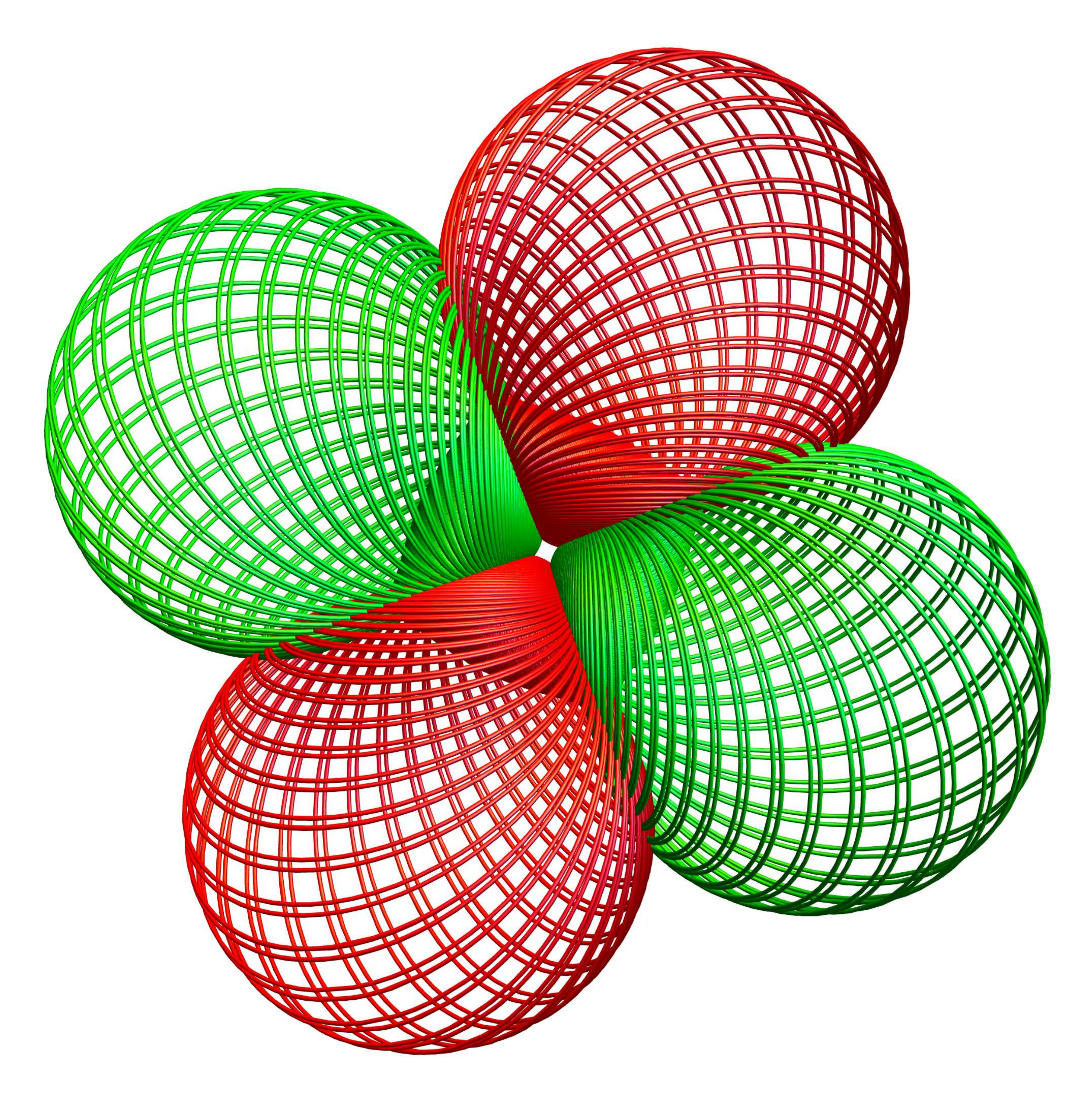}
   \includegraphics[width = 9cm, height = 5cm, trim = 9cm 6.5cm 0cm 3cm, clip]{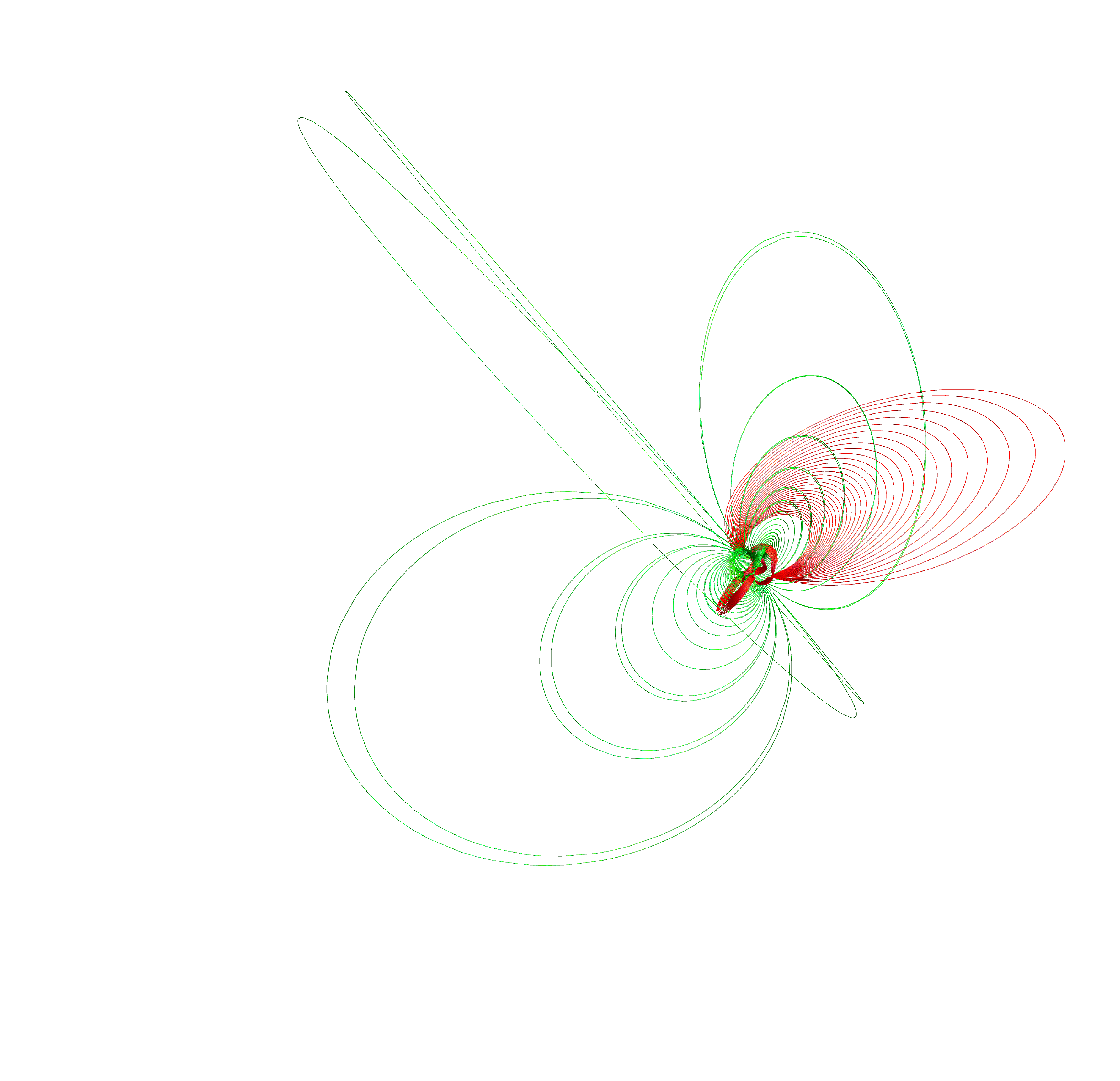}
 \caption{
 Sample electric (red) and magnetic (green) field lines at $t{=}0$.
 Left: $(j;m,n)=(0;0,-1)$ configuration, Centre: $(j;m,n)=(\sfrac12;\sfrac12,\sfrac12)$ configuration, Right: $(j;m,n)=(1;0,1)$ configuration. More self-knotted field lines start appearing with additional seed points in the simulation; the size of these new knots increases with the increase in the seed point's distance from the origin.}
\label{fig2}
\end{figure}
It may be noted here that, due to the compactness of the Lorentzian cylinder, the sphere-frame energy and action are always finite. One can obtain, by a cumbersome calculation, the electric/magnetic fields in Minkowski space by using the results \eqref{1formMap} in \eqref{2FormEM} and realizing that
\begin{equation}\label{2FormMink}
    \Fcal \= F \= E_a\, \diff x^a\wedge \diff t + \sfrac12\, B_a\, \varepsilon^a_{\ bc}\, \diff x^b\wedge \diff x^c\ .
\end{equation}
The actual expression for electric/magnetic fields are rather involved and it bodes well to combine them as the Riemann--Silberstein vector $\textbf{S}$:
\begin{equation}\label{RSvector}
    \textbf{S} \= \textbf{E}+\im \textbf{B}\ .
\end{equation}
The expression of $\textbf{S}$ becomes more and more complicated with increasing $j$.\footnote{Expression of the RS vector for $(j;m,n)=(1;0,0)$ has been recorded in \cite{KL20}.} The (self) knotted field lines corresponding to such electromagnetic configurations also increase in complexity with $j$. We illustrate this point by producing field lines for $j=0,1/2$ and $1$ in Figure \ref{fig2}. 

\section{Conformal group and Noether charges}
\label{sec3}
\noindent
It is well known \cite{BR16} that free Maxwell theory on $\R^{1,3}$ arising from the action
\begin{equation}
    S\left[A_\mu\right] \= \int d^4x\, \mathcal{L}\ ;\qquad \mathcal{L} \= -\sfrac14 F^{\mu\nu}F_{\mu\nu}
\end{equation}
is invariant under the conformal group $SO(2,4)$. Furthermore, the above action is also invariant under the gauge transformations: $A_\mu(x)\rightarrow A_\mu(x)+\pa_\mu\lambda(x)$ for some scalar field $\lambda(x)$. The conformal group is generated by transformations $x^\mu \rightarrow x^\mu + \xi^\mu(x)$, where the vector fields $\xi^\mu$ obey the conformal Killing equations:
\begin{equation}\label{Killing}
    \xi_{\mu,\nu}\, +\, \xi_{\nu,\mu} \= \sfrac12\, \eta_{\mu\nu}\, \xi^\alpha_{~,\alpha} \with \{\eta_{\mu\nu}\} \= \textrm{diag}(-1,1,1,1)\ .
\end{equation}
The conserved Noether current $J^\mu$ is obtained by equating the ``on-shell variation" of the action where the variations $\delta A_\mu$ are arbitrary and the fields $A_\mu$ satisfies the Euler--Langrange equations with its ``symmetry variation" where the fields $A_\mu$ are arbitrary but variations $\delta A_\mu$ satisfy the symmetry condition. The correct variation $\delta A_\mu$ is obtained by imposing the gauge invariance on the Lie derivative of $A_\mu$ w.r.t. the vector field $\xi^\mu$:
\begin{equation}
    \mathcal{L}_{\xi^\alpha}A_\mu := A'_\mu (x) - A_\mu (x) = -\xi^\alpha\, \pa_\alpha A_\mu - A_\alpha\, \pa_\mu \xi^\alpha ~\longrightarrow ~ \delta A_\mu \= F_{\mu\nu}\xi^\nu\ .
\end{equation}
Finally, the conserved current is obtained as
\begin{equation}\label{currentJ}
    J^\mu \= \frac{\pa \mathcal{L}}{\pa A_{\rho,\mu}} \delta A_\rho + \xi^\mu \mathcal{L} \= \xi^\rho \left( F^{\mu\alpha}F_{\rho\alpha} - \sfrac14\delta^\mu_\rho F^2   \right) \with F^2 = F_{\beta \gamma}F^{\beta \gamma}\ ,
\end{equation}
which satisfies the continuity equation and gives the conserved (in time) charge $Q$\footnote{For source-free fields, the current $\textbf{J}$ is assumed to vanish when the surface $\pa V$ is taken to infinity.}:
\begin{equation}
    \pa_\mu\, J^\mu = 0 \quad\implies\quad \frac{\diff Q}{\diff t} \= -\int_{\pa V} \diff^2\textbf{s}\cdot\textbf{J} \= 0\ ;\quad Q \= \int_{V} \diff^3x\, J^0\ .
\end{equation}
Before proceeding further, a couple of remarks pertaining to the subsequent calculations are in order:
\begin{itemize}
   \item All the charges $Q$ are computed at $t\!=0$ owing to the simple $J^0$ expressions on this time-slice. To that end, we record the following useful identities at $t=\tau=0$:
   \begin{equation}\label{t0Results}
      \begin{aligned}
       e^a_i &\= \sfrac1\ell \left( \gamma\, \omega_4\, \delta^a_i - \omega_a\, \omega_i + \epsilon_{aic}\, \gamma\, \omega_c \right)\ ,\quad  e^a_i\, e^b_i \= \sfrac{\gamma^2}{\ell^2} \delta^{ab} \\
       \gamma &\= 1-\omega_4\ ,~ \diff^3x \= \sfrac{\ell^3}{\gamma^3}\diff^3\Omega_3\ ~;~ \diff^3\Omega_3 := e^1\wedge e^2\wedge e^3\ .
      \end{aligned}
   \end{equation}
   Moreover, the electromagnetic fields at $t\!=0$ are given in terms of tetrads $e^\tau = e^\tau_\mu\, \diff x^\mu ,\ e^a = e^a_\mu\, \diff x^\mu$:
   \begin{equation}\label{EMfields}
    E_i \= e^\tau_0\, e^a_i\, \mathcal{E}_a \quad\und\quad  B_i \= \sfrac12\, \epsilon_{ijk}\, \epsilon_{abc}\, e^b_j\, e^c_k\, \mathcal{B}_a\ 
   \end{equation}
   using the two-form \eqref{2FormEM} and its Minkowski counterpart \eqref{2FormMink}.

   \item  Furthermore, these charges $Q$ are computed for a fixed spin-$j$ and, thus, we will suppress the index $j$ from now onwards, unless necessary. Note that the sphere-frame EM fields for fixed $j$ can be obtained by using the expansion \eqref{AviaZ} in \eqref{curFields} as
   \begin{equation} \label{EBj}
     \Acal_a \= Z_a(\omega)\,\ep^{\Omega\,\im\tau} + \bar{Z}_a(\omega)\,\ep^{-\Omega\,\im\tau}
     \qquad\Rightarrow\qquad \biggl\{ \begin{array}{l}
     \Ecal_a \= -\im\,\Omega\,Z_a\,\ep^{\Omega\,\im\tau} + \im\,\Omega\,\bar{Z}_a\,\ep^{-\Omega\,\im\tau} \\[2pt]
     \Bcal_a \= \,-\,\Omega\,Z_a\,\ep^{\Omega\,\im\tau}\,-\,\Omega\,\bar{Z}_a\,\ep^{-\Omega\,\im\tau} \end{array} \biggr\}
   \end{equation}
   with $\bar{Z}_a$ denoting the complex conjugate of $Z_a$.
 
   \item  The simplifications below for the charge density $J^0$ are carried out using the harmonic expansion \eqref{Zharmonics} while using the results \eqref{t0Results} in \eqref{EMfields}.
   
   \item We frequently use below the well-known fact that an odd $\{\omega_{_A}\}$ integral over $S^3$ vanishes because of the opposite contributions coming from the antipodal points on the sphere. In particular, it can be checked that the following integral vanishes\footnote{Note that the power of $\omega_{_A}$ in $Y_{j;m,n}\bar{Y}_{j;m',n'}$ is always even. One way to check this is by employing the toroidal coordinates: $\omega_1 = \cos\eta\cos\zeta_1,\omega_2 = \cos\eta\sin\zeta_1, \omega_3 = \sin\eta\cos\zeta_2,\omega_4 = \sin\eta\sin\zeta_2$ with $\eta\in(0,\sfrac\pi2)$ and $\zeta_1,\zeta_2\in(0,2\pi)$ in \eqref{S3Harmonics}. The resultant selection rules coming from $\zeta_1,\zeta_2$ integral would yield $m-m'\in \sfrac{2k+1}{2}$ with $k\in\N_0$, which is not feasible for fixed $j$.}:
   \begin{equation}
      \int \diff^3\Omega_3\, (\omega_1)^a\, (\omega_2)^b\, (\omega_3)^c\, (\omega_4)^d\, Y_{j;m,n}\, \overline{Y}_{j;m',n'} \= 0 \quad\textrm{for}\quad a+b+c+d\ \in\ 2\N_0 + 1\ .
   \end{equation}
\end{itemize}
Having made these remarks, we now proceed to compute the charges $Q$ for various conformal transformations $\xi^\mu$ obeying \eqref{Killing} in following four categories.

\subsection{Translations}
\noindent
An easily seen solution to \eqref{Killing} is the set of four constant translations
\begin{equation}
    \xi^\mu = \epsilon^\mu\ ,
\end{equation}
which also partly generates the Poincar\'e group and give rise to the usual stress-energy tensor of electrodynamics
\begin{equation}\label{stress-energy}
    J^\mu \= T^\mu_{\ \; \nu} \=  F^{\mu\alpha}F_{\nu\alpha} - \sfrac14\delta^\mu_\nu F^2\ ,
\end{equation}
corresponding to the $\mu$-component of the translation for an arbitrary $\epsilon^\nu$. The corresponding charges are the energy $E$ and the momentum $\textbf{P}$.

\noindent
\textbf{Energy.} The expression of the energy density $e:=T^{00}$ simplifies to  
\begin{equation}
    e\, :=\, \sfrac12 \left( E_i^2 + B_i^2 \right) \= (\sfrac\gamma\ell)^4\, \rho \quad\with\quad \rho \= \sfrac12 \left( \Ecal_a^2 + \Bcal_a^2 \right)\ ,
\end{equation}
which, in turn, simplifies the expression for the energy $E$ to
\begin{equation}
    E \= \int_V \diff^3x\; e \= \sfrac{2}{\ell}\, \Omega^2 \int_{S^3} \diff^3\Omega_3\,  (1-\omega_4)\, Z_a\, \bar{Z}_a\ .
\end{equation}
The resultant expression for $E$ in terms of complex parameters $\Lambda_{m,n}$ and its conjugate was already computed in \cite{KL20}. The $\omega_4$ part of the above integral vanishes as discussed in an earlier remark and we obtain
\begin{equation}
    E \= \frac{8}{\ell}\, (j+1)^3(2j+1)\, \sum_{m,n}\,|\Lambda_{m,n}|^2\ .
\end{equation}

\noindent
\textbf{Momentum.} For the momentum densities $p_i = T^{0i} = (\textbf{E}\times\textbf{B})_i$ we obtain an interesting correspondence relating the one-form $p := p_i\, \diff x^i$ on Minkowski space with a similar one on de Sitter space:
\begin{equation}\label{1formP}
    p \= (\sfrac\gamma\ell)^3\, \mathcal{P}_a\, e^a\ =:\ (\sfrac\gamma\ell)^3\, \mathcal{P} \quad\with\quad \mathcal{P}_a\ :=\ \varepsilon_{abc}\, \mathcal{E}_b\, \mathcal{B}_c\ .
\end{equation}
A straightforward calculation then yields the expression of momenta $P_i$:
\begin{equation}
    P_i \= \int_V \diff^3x\, p_i \= \int_{S^3} \diff^3\Omega_3\, \mathcal{P}_a\, e^a_i \= 2\im\ \Omega^2\ \varepsilon_{abc}\ \int_{S^3} \diff^3\Omega_3\, e^a_i\, Z_b\, \bar{Z}_c
\end{equation}
with $e^a_i$ given by \eqref{t0Results}. The results for $j=0$ are
\begin{equation}
  \begin{aligned}
    P_1^{(j=0)} &\= -\sfrac{\sqrt{2}}{\ell}\left(\left(\bar{\Lambda}_{0,-1} + \bar{\Lambda}_{0,1}\right) \Lambda_{0,0} + \bar{\Lambda}_{0,0}\left(\Lambda_{0,-1}+\Lambda_{0,1}\right)\right)\ , \\
    P_2^{(j=0)} &\=  \sfrac{\im\sqrt{2}}{\ell} \left(\left(-\bar{\Lambda}_{0,-1} + \bar{\Lambda}_{0,1}\right) \Lambda_{0,0} + \bar{\Lambda}_{0,0}\left(\Lambda_{0,-1}-\Lambda_{0,1}\right)\right)\ , \\
    P_3^{(j=0)} &\= \sfrac2\ell \left( |\Lambda_{0,-1}|^2 - |\Lambda_{0,1}|^2 \right)\ .
  \end{aligned}
\end{equation}
As a consistency requirement, we check that the vector charges $P_i$ are rotated according to the algebra of $\mathcal{D}_a$ \eqref{so3rot} (see appendix \ref{appendixA}):
\begin{equation}\label{rotP}
    \mathcal{D}_a\, P_b \= 2\,\varepsilon_{abc}\, P_c\ .
\end{equation}
We also note down $P_3$ for $j=1/2\ ~\textrm{and}~ 1$ in Table \ref{table1}.
\begin{table}[]
    \centering\setcellgapes{4pt}\makegapedcells \renewcommand\theadfont{\normalsize\bfseries}
    \begin{tabular}{|c|P{6cm}|P{8cm}|}
        \hline
      & $j=1/2$ & $j=1$ \\ [0.5ex] \hline \hline
     $P_3$ & 
     $\begin{aligned}
         \sfrac{9}{\ell} \Big( &|\Lambda_{-1/2,-3/2}|^{2} - |\Lambda_{-1/2, 1/2}|^{2} \\
         &- 2 |\Lambda_{-1/2, 3/2}|^{2} + 2 |\Lambda_{1/2,-3/2}|^{2} \\
         &+ |\Lambda_{1/2,-1/2}|^{2} - |\Lambda_{1/2, 3/2}|^{2}\Big)
     \end{aligned}$ & 
     $\begin{aligned}
         \sfrac{24}{\ell}\Big( &|\Lambda_{-1,-2}|^{2} - |\Lambda_{-1,0}|^{2} - 2|\Lambda_{-1,1}|^{2} - 3|\Lambda_{-1,2}|^{2} \\
         &+ 2|\Lambda_{0,-2}|^{2} + |\Lambda_{0,-1}|^{2} - |\Lambda_{0, 1}|^{2} - 2|\Lambda_{0,2}|^{2} \\
         &+ 3|\Lambda_{1,-2}|^{2} + 2|\Lambda_{1,-1}|^{2} + |\Lambda_{1,0}|^{2} - |\Lambda_{1,2}|^{2} \Big)
     \end{aligned}$  \\ \hline
     $P_\phi$ & 
     $\begin{aligned}
         9 \Big( &2|\Lambda_{-1/2,-3/2}|^{2} + |\Lambda_{-1/2,-1/2}|^{2} \\
         &-|\Lambda_{-1/2,3/2}|^{2} + |\Lambda_{1/2,-3/2}|^{2} \\
         &- |\Lambda_{1/2,1/2}|^{2} - 2|\Lambda_{1/2, 3/2}|^{2}\Big)
     \end{aligned}$ & 
     $\begin{aligned}
         24 \Big( &3|\Lambda_{-1,-2}|^{2} + 2|\Lambda_{-1,-1}|^{2} + |\Lambda_{-1,0}|^{2} - |\Lambda_{-1,2}|^{2} \\
         &+ 2|\Lambda_{0,-2}|^{2} + |\Lambda_{0,-1}|^{2} - |\Lambda_{0,1}|^{2} - 2|\Lambda_{0,2}|^{2} \\
         &+ |\Lambda_{1,-2}|^{2} - |\Lambda_{1,0}|^{2} - 2|\Lambda_{1,1}|^{2} - 3|\Lambda_{1,2}|^{2} \Big)
     \end{aligned}$  \\ \hline
     $L_3$ & 
     $\begin{aligned}
         9\ell\, \Big( &-2|\Lambda_{-1/2,-3/2}|^{2} - |\Lambda_{-1/2,-1/2}|^{2} \\
         &+|\Lambda_{-1/2, 3/2}|^{2} - |\Lambda_{1/2,-3/2}|^{2} \\
         &+ |\Lambda_{1/2,1/2}|^{2} + 2|\Lambda_{1/2, 3/2}|^{2}\Big)
     \end{aligned}$ & 
     $\begin{aligned}
         -24\ell\, \Big( &3|\Lambda_{-1,-2}|^{2} + 2|\Lambda_{-1,-1}|^{2} + |\Lambda_{-1,0}|^{2} - |\Lambda_{-1,2}|^{2} \\
         &+ 2|\Lambda_{0,-2}|^{2} + |\Lambda_{0,-1}|^{2} - |\Lambda_{0, 1}|^{2} - 2|\Lambda_{0,2}|^{2} \\
         &+ |\Lambda_{1,-2}|^{2} - |\Lambda_{1,0}|^{2} - 2|\Lambda_{1,1}|^{2} - 3|\Lambda_{1,2}|^{2} \Big)
     \end{aligned}$  \\ \hline
    \end{tabular}
    \caption{Expressions of $P_3$, $P_
    \phi$ and $L_3$ for $j=1/2 ~\textrm{and}~ 1$.}
    \label{table1}
\end{table}
One can compute the corresponding $P_1$ and $P_2$ for $j=1/2$ and $j=1$ by employing the action of an appropriate $\mathcal{D}_a$ of Table \ref{table4}.

We can additionally compute the spherical components of the momentum $(P_r,P_\theta,P_\phi)$ by letting the one-form $e^a$ in \eqref{1formP} act on the vector fields $(\pa_r,\pa_\theta, \pa_\phi)$. In practice, we first write $\pa_r = -\sfrac\gamma\ell \pa_\chi$ using \eqref{Jacobian} at $t=0$ and then invert the vector fields $(\pa_r,\pa_\theta, \pa_\phi)$ in terms of the left invariant vector fields $(L_1,L_2,L_3)$ using \eqref{Lfields}. Finally, using the duality relation $e^a(L_b)=\delta^a_b$ we obtain
\begin{equation}\label{sphericalP}
 \begin{aligned}
   P_r &\= -\frac1\ell\int_{S^3} \diff^3\Omega_3\,(1-\cos\chi)\, \left( \sin\theta\cos\phi\, \mathcal{P}_1 + \sin\theta\sin\phi\, \mathcal{P}_2 + \cos\theta\, \mathcal{P}_3 \right)\ , \\
   P_\theta &\= \int_{S^3} \diff^3\Omega_3\,\sin\chi\cos\chi \left( (\cos\theta\cos\phi + \tan\chi\sin\phi)\, \mathcal{P}_1 + (\cos\theta\sin\phi - \tan\chi\cos\phi)\mathcal{P}_2 - \sin\theta\, \mathcal{P}_3 \right) ,\\
   P_\phi &\= \int_{S^3} \diff^3\Omega_3\, \sin^2\chi\sin\theta \left( (\cos\theta\cos\phi - \cot\chi\sin\phi)\, \mathcal{P}_1 + (\cos\theta\sin\phi + \cot\chi\cos\phi)\, \mathcal{P}_2 - \sin\theta\, \mathcal{P}_3 \right) \ .
 \end{aligned}
\end{equation}
From the expression of $P_r$ we see that the integrand over $S^2$, i.e. $\hat{\omega}_a\mathcal{P}_a$ \eqref{omegas} is an odd function,\footnote{The terms of $\mathcal{P}_a$ using \eqref{adjointY} are proportional to  $Y_{l,M}Y_{l',M'}$, which is even in $\hat{\omega}_a$ for a fixed $j$.} which makes $P_r$ vanish. We also find with explicit calculations (verified for up to $j=1$) that $P_\theta$ vanishes. For $j=0$ we find that $P_\phi$ are proportional to $P_3$:
\begin{equation}
    P_\phi^{(j=0)} \= \ell\, P_3^{(j=0)} \=  2\left(|\Lambda_{0,-1}|^{2}-|\Lambda_{0,1}|^{2}\right)\ .
\end{equation}
The expressions of $P_\phi$ for $j=1/2$ and $1$ has been recorded in the Table \ref{table1}.
\subsection{Lorentz transformations}
\noindent
Another solution of \eqref{Killing} is given by
\begin{equation}
    \xi^\mu(x) \= \epsilon^{\mu}_{\ \; \nu}\, x^\nu \quad\textrm{where}\quad \epsilon_{\mu\nu} \= -\epsilon_{\nu\mu}\ ,
\end{equation}
which correspond to the six generators of the Lorentz group $SO(1,3)$. These six together with the above four translations generates the full Poincar\'e group. The corresponding six charges for these transformations are grouped into the boost $\textbf{K}$ and the angular momentum $\textbf{L}$.

\noindent
\textbf{Boost.} The conserved charge densities arising from $J^0$ \eqref{currentJ} corresponding to $\epsilon_{0i}$ are the boost densities $\textbf{k} = \rho\ \textbf{x} - \textbf{p}\ t$, which simplify for $t=0$ to
\begin{equation}
    k_i \= (\sfrac\gamma\ell)^3\, \mathcal{K}_i \quad\with\quad \mathcal{K}_i \= \rho\, \omega_i\ .
\end{equation}
The corresponding charges $K_i$ vanish because of the odd integrand as discussed in an earlier remark:
\begin{equation}
K_i \= \int_V \diff^3x\, k_i \= \int_{S^3} \diff^3\Omega_3\, \mathcal{K}_i \= 2 \int_{S^3} \diff^3\Omega_3\, \omega_i\, Z_a^j\, \bar{Z}_a^j \= 0\ .
\end{equation}

\noindent
\textbf{Angular momentum.} The other three conserved charge densities $J^0$ corresponding to $\epsilon_{ij}$ in \eqref{currentJ} are the angular momentum densities $\textbf{l} = \textbf{p}\times\textbf{x}$, which take a simple form just like in momentum \eqref{1formP}:
\begin{equation}
    l_i\, \diff x^i \= (\sfrac\gamma\ell)^2\, \mathcal{L}_a\, e^a \quad\with\quad \mathcal{L}_a \= \varepsilon_{abc}\, \mathcal{P}_b\, \omega_c\ .
\end{equation}
The expressions for the charges $L_i$ simplify to
\begin{equation}
    L_i \= \int_V \diff^3x\, l_i \= \int_{S^3} \diff^3\Omega_3\, (\sfrac\ell\gamma)\, \mathcal{L}_a\, e^a_i \= 2\im\ell\ \Omega^2\ \int_{S^3} \diff^3\Omega_3\, \sfrac{1}{\gamma}\, e^a_i \left( \bar{Z}_a\, \omega_b Z_b - Z_a\, \omega_b\bar{Z}_b \right)\ .
\end{equation}
Explicit calculations show that for $j=0$ the angular momenta $L_i$ are proportional to the momenta $P_i$:
\begin{equation}
    L_i^{(j=0)} \= -\ell^2\, P_i^{(j=0)}\ .
\end{equation}
This is, however, not true for higher spin $j$. The angular momenta $L_i$ has the same rotation behaviour as for the momenta $P_i$ \eqref{rotP}. We, therefore, note down the results of $L_3$ for $j=1/2$ and $1$ in Table \ref{table1}, from which the corresponding expressions of $L_1$ and $L_2$ can be obtained using Table \ref{table4}.

We can again compute the spherical components of the angular momentum $(L_r,L_\theta,L_\phi)$ using the relations \eqref{sphericalP} by replacing $\mathcal{P}$ with $\mathcal{L}$ in it. We realize that the $S^2$ integrand for $L_r$ would only have terms like $\hat{\omega}_a\hat{\omega}_b\mathcal{P}_c$, which are all odd functions\footnote{Here again the terms of $\mathcal{P}_c$, which goes like $Y_{l,M}Y_{l',M'}$ \eqref{adjointY}, are all even functions of $\phi$ for a fixed $j$.} of $\phi$ and, therefore, the $\phi$ integral over the domain $(0,2\pi)$ would make $L_r$ vanish. We also find, with explicit computations, that the charges $L_\theta$ and $L_\phi$ for $j=0$ are proportional to $P_3$:
\begin{equation}
    L_\theta^{(j=0)} \= \sfrac43\, \ell^2\, P_3^{(j=0)} \quad\und\quad L_\phi^{(j=0)} \= -\sfrac13\, \ell^2\, P_3^{(j=0)}\ .
\end{equation}
As a non trivial example, we collect these charges for $j=1/2$ below:
\begin{equation}
  \begin{aligned}
     L_\theta^{\left(j=1/2\right)} \= \sfrac{12}{5}\ell\, \Big( &9|\Lambda_{-1/2,-3/2}|^{2} + 6|\Lambda_{-1/2,-1/2}|^{2} +|\Lambda_{-1/2, 1/2}|^{2}
     - 6|\Lambda_{-1/2,3/2}|^{2} \\ 
     &\qquad\qquad\qquad\qquad + 6|\Lambda_{1/2,-3/2}|^{2} - |\Lambda_{1/2, -1/2}|^{2} - 6|\Lambda_{1/2,1/2}|^{2} - 9|\Lambda_{1/2, 3/2}|^{2} \Big)\ ,
  \end{aligned}
\end{equation}
\begin{equation}
  \begin{aligned}
      L_\phi^{\left(j=1/2\right)} \= -\sfrac35 \ell\, \Big(  &6|\Lambda_{-1/2,-3/2}|^{2} - |\Lambda_{-1/2,-1/2}|^{2} -6|\Lambda_{-1/2, 1/2}|^{2} - 9|\Lambda_{-1/2,3/2}|^{2} + 9|\Lambda_{1/2,-3/2}|^{2} \\
      &\qquad\quad + 6|\Lambda_{1/2,-1/2}|^{2} + |\Lambda_{1/2,1/2}|^{2} - 6|\Lambda_{1/2, 3/2}|^{2} + \sqrt{3}\big( \bar{\Lambda}_{1/2,-3/2}\Lambda_{-1/2,-1/2}\\ 
      &\qquad\qquad\qquad -\bar{\Lambda}_{1/2,1/2}\Lambda_{-1/2,3/2} + \bar{\Lambda}_{-1/2,-1/2}\Lambda_{1/2,-3/2} - \bar{\Lambda}_{-1/2,3/2}\Lambda_{1/2,1/2}\big) \Big)\ .
  \end{aligned}
\end{equation}

\subsection{Dilatation} 
It is easy to verify that a constant rescaling by $\lambda$:
\begin{equation}
    \xi^\mu \= \lambda\, x^\mu
\end{equation}
is also a solution of \eqref{Killing}. The charge density corresponding to this single generator of the conformal group is $\textbf{p}\cdot\textbf{x}- e\, t$, which for $t\!=0$ simplifies to 
\begin{equation}
    p_i\, x_i \= (\sfrac\gamma\ell)^3\, \mathcal{P}_a\,\omega_a\ .
\end{equation}
The corresponding charge $D$ vanishes because of the odd integrand:
\begin{equation}
    D \= \int_V \diff^3x\, p_i\, x_i \= 2\im\,\Omega^2\, \varepsilon_{abc} \int_{S^3} \diff^3\Omega_3\, \omega_a\, Z_b\, \bar{Z}_c \= 0\ .
\end{equation}
\subsection{Special conformal transformations}
A fairly straightforward calculation shows that the following not so obvious transformation
\begin{equation}
    \xi^\mu \= 2\, x^\mu\,b_\nu x^\nu\ -\ b^\mu\, x^\nu x_\nu 
\end{equation}
also satisfies \eqref{Killing}. The four generators corresponding to $b_\mu$ give rise to four different charges $V_0$ and $\textbf{V}$.

\noindent
\textbf{Scalar SCT.} The charge density $J^0$ corresponding to $b_0$ is $v_0 = (\textbf{x}^2+t^2)\, e - 2t\, \textbf{p}\cdot\textbf{x}$, which for $t\!=0$ simplifies to
\begin{equation}
    v_0 \= \textbf{x}^2\, e \= (\sfrac\gamma\ell)^2 \rho\, \omega_a^2\ .
\end{equation}
The expression for the corresponding charge $V_0$ takes the following simple form:
\begin{equation}
    V_0 \= \int_V \diff^3x\, v_0 \= \int_{S^3} \diff^3\Omega_3\, (\sfrac\ell\gamma)\, (1-\omega_4^2)\, \rho \= 2\ell\, \Omega^2 \int_{S^3} \diff^3\Omega_3\, (1+\omega_4)\, Z_a\, \bar{Z}_a\ .
\end{equation}
Here again the $\omega_4$ term of the integral, being odd, vanishes and yields
\begin{equation}
    V_0 \= \ell^2\, E \= 8\,\ell\, (j+1)^3(2j+1)\, \sum_{m,n}\,|\Lambda_{m,n}|^2\ .
\end{equation}

\noindent
\textbf{Vector SCT.} The charge densities $J^0$ that correspond to $b_i$ read: $\textbf{v} = 2\textbf{x}(\textbf{x}\cdot\textbf{p}) - 2t\,\textbf{x}\,e - (\textbf{x}^2-t^2)\textbf{p}$. This simplify at $t\!=0$ and take a structure similar to the momentum densities $p_i$ \eqref{1formP}:
\begin{equation}\label{vectorSCT}
    v_i\, dx^i \= (\sfrac\gamma\ell)\, \mathcal{V}_a\, e^a \with \mathcal{V}_a \= 2\omega_a\, (\mathcal{P}_b\,\omega_b) - \omega_b^2\,\mathcal{P}_a\ . 
\end{equation}
The expressions for the charges $V_i$ then simplify to
\begin{equation}
    V_i \= \int_V \diff^3x\, v_i \= 2\im\,\ell^2\,\Omega^2 \int_{S^3} \diff^3\Omega_3\, \gamma^{-2} e^a_i \left( 2\varepsilon_{bcd}\,\omega_a\,\omega_b\,Z_c\,\bar{Z}_d - \varepsilon_{abc}\,(1-\omega_4^2)\,Z_b\,\bar{Z}_c \right)\ .
\end{equation}
With explicit computation we observe that the charges $V_i$ are proportional to the momenta $P_i$ (verified explicitly for up to $j=1$)
\begin{equation}
    V_i \= \ell^2\, P_i\ .
\end{equation}

As before, we can compute the spherical components $(V_r,V_\theta,V_\phi)$ by using the expressions \eqref{sphericalP} and replacing $\mathcal{P}$ with $\mathcal{V}$ in it. We notice that the charge $V_r$ vanishes owing to the odd $S^2$ integrand\footnote{Observe that the terms in $\mathcal{V}_a$ \eqref{vectorSCT} are all even in $\hat{\omega}_a$.} just like in the case of $P_r$. However, unlike $P_\theta$ here the charge $V_\theta$ is non-vanishing. Explicit calculations show that for $j=0$ the charges $V_\theta$ and $V_\phi$ are proportional to the momentum $P_3$:
\begin{equation}
    V_\theta^{(j=0)} \= -\sfrac43 \ell^3\, P_3^{(j=0)} \quad\und\quad V_\phi^{(j=0)} \= -\sfrac53 \ell^3\, P_3^{(j=0)}\ .
\end{equation}
Additionally, we record below the charges $V_\theta$ and $V_\phi$ for the non-trivial case of $j=1/2$
\begin{equation}
  \begin{aligned}
     V_\theta^{\left(j=1/2\right)} \= -\sfrac{6}{5}\ell^2\, \Big( &9|\Lambda_{-1/2,-3/2}|^{2} + |\Lambda_{-1/2,-1/2}|^{2} - 9|\Lambda_{-1/2,1/2}|^{2} - 21|\Lambda_{-1/2,3/2}|^{2} \\ 
     &\qquad\qquad\qquad\qquad + 21|\Lambda_{1/2,-3/2}|^{2} +
     9|\Lambda_{1/2,-1/2}|^{2} - |\Lambda_{1/2,1/2}|^{2} - 9|\Lambda_{1/2, 3/2}|^{2} \Big)\ ,
  \end{aligned}
\end{equation}
\begin{equation}
  \begin{aligned}
      V_\phi^{\left(j=1/2\right)} \= -\sfrac35 \ell^2\, \Big(  &42|\Lambda_{-1/2,-3/2}|^{2} + 33|\Lambda_{-1/2,-1/2}|^{2} + 8|\Lambda_{-1/2,1/2}|^{2} - 33|\Lambda_{-1/2,3/2}|^{2} + 33|\Lambda_{1/2,-3/2}|^{2} \\
      &\qquad - 8|\Lambda_{1/2,-1/2}|^{2} - 33|\Lambda_{1/2,1/2}|^{2} - 42|\Lambda_{1/2,3/2}|^{2} + 8\sqrt{3}\big( -\bar{\Lambda}_{1/2,-3/2}\Lambda_{-1/2,-1/2}\\ 
      &\qquad\qquad\quad\quad +\bar{\Lambda}_{1/2,1/2}\Lambda_{-1/2,3/2} - \bar{\Lambda}_{-1/2,-1/2}\Lambda_{1/2,-3/2} + \bar{\Lambda}_{-1/2,3/2}\Lambda_{1/2,1/2}\big) \Big)\ .
  \end{aligned}
\end{equation}
One can compute these charges for higher spin $j$ by following the same strategy.

\section{Applications}
\label{sec4}
\noindent
The method of constructing rational electromagnetic fields presented in this paper has the added advantage that it produces a complete set labelled by $(j,m,n)$; any electromagnetic field configuration having finite energy can, in principle, be obtained from an expansion like in (\ref{AviaZ}-\ref{basisSoln}), albeit with a varying $j$. The operational difficulty involved in this procedure has to do with the fact that this set is infinite as $j\in \sfrac{\N}{2}$. There are, however, many important cases where only a finite number of knot-basis solutions (sometimes only with a fixed $j$) need to be combined to get the desired EM field configuration. One such very important case is that of Hopfian solution given by Ran\~{a}da, which was already covered in \cite{LZ18}. Below we analyse two very interesting generalizations of Hopfian solution presented in \cite{HSS15} in the context of present construction. It is imperative to note here that while the scope of construction of a new solution from the known ones as presented in \cite{HSS15} is limited, the same is not true for the method presented in this paper, which by design can produce arbitrary number of new field configurations. Some of these possible new field configurations obtained from the $j\!=\!0$ sector (possibly from $j\!=\!1/2~ \textrm{or}~ 1$ as well) could find experimental application like, e.g. in \cite{LSMetal18}. 

\begin{figure}[h!]
\centering
   \includegraphics[width = 5cm, height = 5cm]{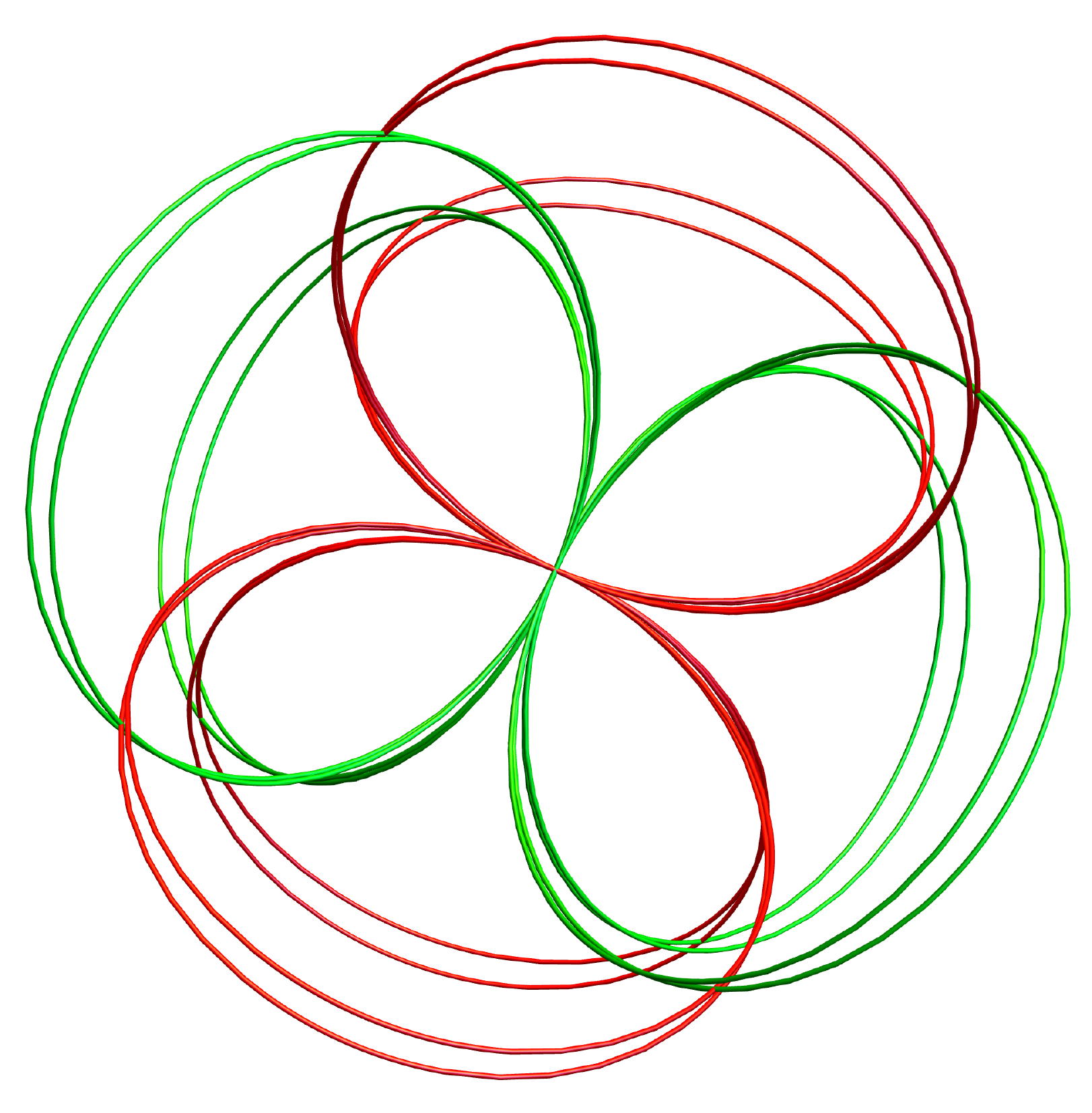}
   \hspace{2cm}
   \includegraphics[width = 5cm, height = 5cm, trim = {2cm 3cm 1.5cm 2cm},clip]{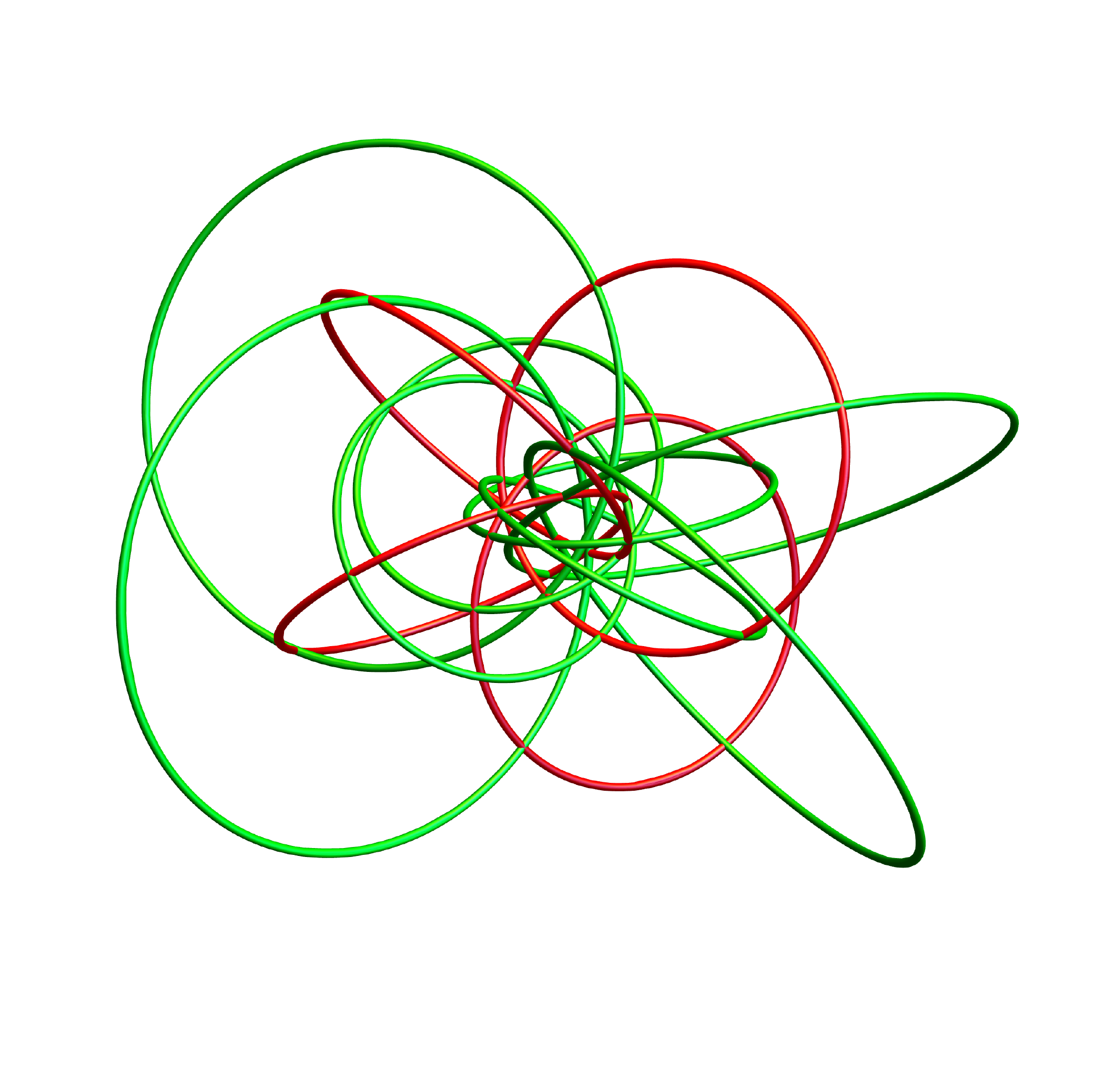}
 \caption{
 Sample electric (red) and magnetic (green) field lines at $t{=}0$.
 Left: time-translated Hopfian with $c=60$, Right: Rotated Hopfian with $\theta=1$.} 
\label{fig3}
\end{figure}
Bateman's construction, employed in \cite{HSS15}, hinges on a judicious ansatz for the Riemann--Silberstein vector \eqref{RSvector} satisfying Maxwell's equations:
\begin{equation}
    \textbf{S} \= \nabla\alpha \times \nabla\beta \quad\implies\quad \nabla\cdot\textbf{S}\=0 ~\ \& ~\ \im\, \partial_t\textbf{S} \= \nabla\times\textbf{S}\ ,
\end{equation}
using a pair of complex functions $(\alpha,\beta)$. An interesting generalization of the Hopfian solution is obtained in equations (3.16-3.17) of \cite{HSS15} using (complex) time-translation (TT) to obtain the following $(\alpha,\beta)$ pair (up to a normalization)
\begin{equation}
    (\alpha,\beta)_{TT} \= \left(\frac{A-1-\im z}{A+\im(t+\im c)}, \frac{x-\im y}{A+\im(t+\im c)} \right)\ ;\quad A\= \sfrac12 \left(x^2+y^2+z^2-(t+\im c)^2+1\right)\ ,
\end{equation}
where $c$ is a constant real parameter. The corresponding  EM field configuration is obtained, in our case, by choosing $\ell = 1-c$ and only the following $j=0$ complex coefficients in \eqref{basisChange}
\begin{equation}\label{j0TT}
    \left(\Lambda_{0;0,1}\ ,~\Lambda_{0;0,0}\ ,~\Lambda_{0;0,-1}\right)_{TT} \= \left( 0\ ,~0\ ,~-\im\frac{\pi}{2\ell^2} \right)\ .
\end{equation}
A sample electromagnetic knot configuration of this modified Hopfian is illustrated in Figure \ref{fig3}. Another interesting generalization of the Hopfian is constructed in equations (3.20-3.21) of \cite{HSS15} using a (complex) rotation (R) to get the following $(\alpha,\beta)$ pair (again, up to a normalization)
\begin{equation}
    (\alpha,\beta)_{R} \= \left(\frac{A-1+\im(z \cos{\im\theta}+x\sin{\im\theta})}{A+\im t}, \frac{x \cos{\im\theta}-z\sin{\im\theta}-\im y}{A+\im t} \right)\ ;\ A\= \sfrac12 \left(x^2+y^2+z^2-t^2+1\right)\ .
\end{equation}
To get this particular EM field configuration, we need to set $\ell=1$ and use the following combination of $j=0$ coefficients in \eqref{basisChange}
\begin{equation}\label{j0R}
    \left(\Lambda_{0;0,1}\ ,~\Lambda_{0;0,0}\ ,~\Lambda_{0;0,-1}\right)_{R} \= \left( \im\frac{\pi}{4}(\cosh{\theta}-1)\ ,~-\frac{\pi}{2\sqrt{2}}\sinh{\theta}\ ,~-\im\frac{\pi}{4}(\cosh\theta+1) \right)\ .
\end{equation}
We illustrate the EM field lines for a sample $\theta$ value of this modified Hopfian in Figure \ref{fig3}. We note down the conformal charges corresponding to these solutions in Table \ref{table3} by plugging the $j=0$ coefficients \eqref{j0TT} and \eqref{j0R} in the appropriate formulae of the previous section. The results match with the ones given in \cite{HSS15} up to a rescaling of the energy, which can be achieved by an appropriate choice of normalization.
\begin{table}[]
    \centering\setcellgapes{4pt}\makegapedcells \renewcommand\theadfont{\normalsize\bfseries}
    \begin{tabular}{|c|P{5cm}|P{5cm}|}
        \hline
      & Time-translated Hopfian & Rotated Hopfian \\ [0.5ex] \hline \hline
     Energy (E) & 
     $\frac{2\pi^2}{(1-c)^5}=:E_{TT}$ & 
     $2\pi^2\cosh^2\theta =: E_{R}$  \\ \hline
     Momentum ($\textbf{P}$) & 
     $\left(0\ ,~0\ ,~\frac{1}{4}\right)E_{TT}$ & 
     $\left(0\ ,~-\frac{1}{4}\tanh{\theta}\ ,~\frac{1}{4}\sech\theta\right)E_{R}$ \\ \hline
     Boost ($\textbf{K}$) & 
     $\left(0\ ,~0\ ,~0\right)$ & 
     $\left(0\ ,~0\ ,~0\right)$ \\ \hline
     Ang. momentum ($\textbf{L}$) & 
     $\left(0\ ,~0\ ,~-\frac{1}{4}(1-c)^2\right)E_{TT}$ & 
     $\left(0\ ,~\frac{1}{4}\tanh{\theta}\ ,~-\frac{1}{4}\sech\theta\right)E_{R}$ \\ \hline
      Dilatation (D) & 
     $0$ & 
     $0$ \\ \hline
     Scalar SCT ($V_0$) & 
     $(1-c)^2E_{TT}$ & 
     $E_{R}$  \\ \hline
     Vector SCT ($\textbf{V}$) & 
     $\left(0\ ,~0\ ,~\frac{1}{4}(1-c)^2\right)E_{TT}$ & 
     $\left(0\ ,~-\frac{1}{4}\tanh{\theta}\ ,~\frac{1}{4}\sech\theta\right)E_{R}$ \\ \hline
    \end{tabular}
    \caption{Conformal charges for the time-translated and rotated Hopfian configurations.}
    \label{table3}
\end{table}

\section*{Acknowledgements}
\noindent
KK is grateful to Deutscher Akademischer Austauschdienst (DAAD) for the doctoral research grant 57381412. We thank Olaf Lechtenfeld for several insightful discussions and valuable suggestions.

\appendix
\section{Rotation of indices}
\label{appendixA}
\noindent
By construction the gauge potential $\Acal$ is $SO(4)$ invariant, which means it is also invariant under the action of $SO(3)$ generators $\mathcal{D}_a$ \eqref{so3rot}. For a complex-valued $\Acal = \Acal_a\,e^a$ expanded using \eqref{AviaZ} and \eqref{Zharmonics} at a fixed $j$ this means 
\begin{equation}
    0\= \mathcal{D}_a (\Acal) \= \sum_{m,n,\tilde{n}} C^{n,\tilde{n}}_{j,b}\, \ep^{\im\Omega\tau}\, \left( \mathcal{D}_a(\Lambda_{m,\tilde{n}})\,e^b\,Y_{j;m,n} + \Lambda_{m,\tilde{n}}\,\mathcal{D}_a(e^b)\,Y_{j;m,n} + \Lambda_{m,\tilde{n}}\,e^b\,\mathcal{D}_a(Y_{j;m,n})  \right)
\end{equation}
where $\mathcal{D}_a(e^b)$ are determined from \eqref{LieAction} while $\mathcal{D}_a(Y_{j;m,n})$ are determined from (\ref{Yalgebra}-\ref{LadderAction}). By collecting the coefficients of various linearly independent $e^b$ and $Y_{j;m,n}$ terms in the above expansion for a fixed $\mathcal{D}_a$ one gets a set of coupled linear equations for $\mathcal{D}_a(\Lambda_{m,\tilde{n}})$, which can be easily solved. The action of the generators $\mathcal{D}_a$ on $\Lambda_{m,\tilde{n}}$ for $j=0,~ 1/2 ~\textrm{and}~ 1$ is given in Table \ref{table4}.
\begin{table}[]
    \resizebox{\columnwidth}{!}{
    \hspace{0cm}\setcellgapes{4pt}\makegapedcells \renewcommand\theadfont{\normalsize\bfseries}
    \begin{tabular}{|P{3.2cm}|P{6cm}|P{6cm}|P{2cm}|}
        \hline
      & $\mathcal{D}_1$ & $\mathcal{D}_2$ & $\mathcal{D}_3$ \\ [0.5ex] \hline \hline
     $\left.j=0: \right.                                                
     \begin{aligned}
        &\Lambda_{0,-1} \mapsto \\
        &\Lambda_{0,0} \mapsto \\
        &\Lambda_{0,1} \mapsto
     \end{aligned}$ & 
     $\begin{aligned}                                               
         &\sqrt{2}\im \Lambda_{0,0} \\
         &\sqrt{2}\im (\Lambda_{0,1}+\Lambda_{0,-1}) \\
         &\sqrt{2}\im \Lambda_{0,0}
     \end{aligned}$ & 
     $\begin{aligned}                                               
         &-\sqrt{2}  \Lambda_{0,0} \\
         &\sqrt{2}  (\Lambda_{0,-1} - \Lambda_{0,1}) \\
         &\sqrt{2}  \Lambda_{0,0} 
     \end{aligned}$ &
     $\begin{aligned}                                               
         &-\sqrt{2}\im  \Lambda_{0,-1} \\
         &\quad 0 \\
         &\sqrt{2}\im  \Lambda_{0,1}
     \end{aligned}$ \\ \hline
     $\left.\begin{aligned}
        &\quad j=\sfrac12: \\                                                
        &\underbrace{\textrm{Notation}}\\
        &\pm\sfrac12\equiv\pm \\
        &\pm\sfrac32\equiv\uparrow\downarrow
     \end{aligned}\right\}
     \begin{aligned}
        &\Lambda_{-,\downarrow} \mapsto \\[.2em]
        &\Lambda_{-,-} \mapsto \\[.2em]
        &\Lambda_{-,+} \mapsto \\[.2em]
        &\Lambda_{-,\uparrow} \mapsto \\[.2em]
        &\Lambda_{+,\downarrow} \mapsto \\[.2em]
        &\Lambda_{+,-} \mapsto \\[.2em]
        &\Lambda_{+,+} \mapsto \\[.2em]
        &\Lambda_{+,\uparrow} \mapsto
     \end{aligned}$ & 
     $\begin{aligned}                                               
         &\im(\sqrt{3}\Lambda_{-,-} + \Lambda_{+,\downarrow} ) \\
         &\im(\sqrt{3}\Lambda_{-,\downarrow} + 2\Lambda_{-,+} + \Lambda_{+,-} ) \\
         &\im(2\Lambda_{-,-} + \sqrt{3}\Lambda_{-,\uparrow} + \Lambda_{+,+} ) \\
         &\im(\sqrt{3}\Lambda_{-,+} + \Lambda_{+,\uparrow} ) \\
         &\im(\Lambda_{-,\downarrow} + \sqrt{3}\Lambda_{+,-} ) \\
         &\im(\Lambda_{-,-} + \sqrt{3}\Lambda_{+,\downarrow} + 2 \Lambda_{+,+} ) \\
         &\im(\Lambda_{-,+} + 2\Lambda_{+,-} + \sqrt{3}\Lambda_{+,\uparrow} ) \\
         &\im(\Lambda_{-,\uparrow} + \sqrt{3}\Lambda_{+,+} ) \\
     \end{aligned}$ &
     $\begin{aligned}                                               
         &-\sqrt{3}\Lambda_{-,-} - \Lambda_{+,\downarrow} \\
         &\sqrt{3}\Lambda_{-,\downarrow} - 2\Lambda_{-,+} - \Lambda_{+,-} \\
         &2\Lambda_{-,-} - \sqrt{3}\Lambda_{-,\uparrow} - \Lambda_{+,+} \\
         &\sqrt{3}\Lambda_{-,+} - \Lambda_{+,\uparrow} \\
         &\Lambda_{-,\downarrow} - \sqrt{3}\Lambda_{+,-} \\
         &\Lambda_{-,-} + \sqrt{3}\Lambda_{+,\downarrow} -2\Lambda_{+,+} \\
         &\Lambda_{-,+} +2\Lambda_{+,-} -\sqrt{3}\Lambda_{+,\uparrow} \\
         &\Lambda_{-,\uparrow} + \sqrt{3}\Lambda_{+,+} \\
     \end{aligned}$ &
     $\begin{aligned}                                               
         &-4\im\Lambda_{-,\downarrow} \\[.2em]
         &-2\im\Lambda_{-,-} \\[.2em]
         &0 \\[.2em]
         &2\im\Lambda_{-,\uparrow} \\[.2em]
         &-2\im\Lambda_{+,\downarrow} \\[.2em]
         &0 \\[.2em]
         &2\im\Lambda_{+,+} \\[.2em]
         &4\im\Lambda_{+,\uparrow}
     \end{aligned}$ \\ \hline
     $\left.\begin{aligned}
        &\quad j=1: \\                                                
        &\underbrace{\textrm{Notation}}\\
        &\pm1\equiv\pm \\
        &\pm2\equiv\uparrow\downarrow
     \end{aligned}\right\}
     \begin{aligned}
        &\Lambda_{-,\downarrow} \mapsto \\[.2em]
        &\Lambda_{-,-} \mapsto \\[.2em]
        &\Lambda_{-,0} \mapsto \\[.2em]
        &\Lambda_{-,+} \mapsto \\[.2em]
        &\Lambda_{-,\uparrow} \mapsto \\[.2em]
        &\Lambda_{0,\downarrow} \mapsto \\[.2em]
        &\Lambda_{0,-} \mapsto \\[.2em]
        &\Lambda_{0,0} \mapsto \\[.2em]
        &\Lambda_{0,+} \mapsto \\[.2em]
        &\Lambda_{0,\uparrow} \mapsto \\[.2em]
        &\Lambda_{+,\downarrow} \mapsto \\[.2em]
        &\Lambda_{+,-} \mapsto \\[.2em]
        &\Lambda_{+,0} \mapsto \\[.2em]
        &\Lambda_{+,+} \mapsto \\[.2em]
        &\Lambda_{+,\uparrow} \mapsto
     \end{aligned}$ & 
     $\begin{aligned}                                               
         &\im(2\Lambda_{-,-} + \sqrt{2}\Lambda_{0,\downarrow} ) \\
         &\im(2\Lambda_{-,\downarrow} + \sqrt{6}\Lambda_{-,0} + \sqrt{2}\Lambda_{0,-} ) \\
         &\im(\sqrt{6}\Lambda_{-,-} + \sqrt{6}\Lambda_{-,+} + \sqrt{2}\Lambda_{0,0} ) \\
         &\im(\sqrt{6}\Lambda_{-,0} + 2\Lambda_{-,\uparrow} + \sqrt{2}\Lambda_{0,+} ) \\
         &\im(2\Lambda_{-,+} + \sqrt{2}\Lambda_{+,\uparrow} ) \\
         &\im(\sqrt{2}\Lambda_{-,\downarrow} + 2\Lambda_{0,-} + \sqrt{2}\Lambda_{+,\downarrow} ) \\
         &\im\sqrt{2}(\Lambda_{-,-} + \sqrt{2}\Lambda_{0,\downarrow} + \sqrt{3}\Lambda_{0,0} + \Lambda_{+,-} ) \\
         &\im\sqrt{2}(\Lambda_{-,0} + \sqrt{3}\Lambda_{0,-} + \sqrt{3}\Lambda_{0,+} + \Lambda_{+,0} ) \\
         &\im\sqrt{2}(\Lambda_{-,+} + \sqrt{3}\Lambda_{0,0} + \sqrt{2}\Lambda_{0,\uparrow} + \Lambda_{+,+} ) \\
         &\im(\sqrt{2}\Lambda_{-,\uparrow} + 2\Lambda_{0,+}  + \sqrt{2}\Lambda_{+,\uparrow} ) \\
         &\im(\sqrt{2}\Lambda_{0,\downarrow} + 2\Lambda_{+,-} ) \\
         &\im(\sqrt{2}\Lambda_{0,-} + 2\Lambda_{+,\downarrow} + \sqrt{6}\Lambda_{+,0} ) \\
         &\im(\sqrt{2}\Lambda_{0,0} + \sqrt{6}\Lambda_{+,-} + \sqrt{6}\Lambda_{+,+} ) \\
         &\im(\sqrt{2}\Lambda_{0,+} + \sqrt{6}\Lambda_{+,0} + 2\Lambda_{+,\uparrow} ) \\
         &\im(\sqrt{2}\Lambda_{0,\uparrow} + 2\Lambda_{+,+} ) \\        
     \end{aligned}$ &
     $\begin{aligned}                                               
         &-2\Lambda_{-,-} - \sqrt{2}\Lambda_{0,\downarrow} \\
         &2\Lambda_{-,\downarrow} - \sqrt{6}\Lambda_{-,0} - \sqrt{2}\Lambda_{0,-} \\
         &\sqrt{6}\Lambda_{-,-} - \sqrt{6}\Lambda_{-,+} - \sqrt{2}\Lambda_{0,0} \\
         &\sqrt{6}\Lambda_{-,0} - 2\Lambda_{-,\uparrow} - \sqrt{2}\Lambda_{0,+} \\
         &2\Lambda_{-,+} - \sqrt{2}\Lambda_{0,\uparrow} \\
         &\sqrt{2}\Lambda_{-,\downarrow} - 2\Lambda_{0,-} - \sqrt{2}\Lambda_{+,\downarrow} \\
         &\sqrt{2}(\Lambda_{-,-} + \sqrt{2}\Lambda_{0,\downarrow} - \sqrt{3}\Lambda_{0,0} - \Lambda_{+,-}) \\
         &\sqrt{2}(\Lambda_{-,0} + \sqrt{3}\Lambda_{0,-} - \sqrt{3}\Lambda_{0,+} - \Lambda_{+,0}) \\
         &\sqrt{2}(\Lambda_{-,+} + \sqrt{3}\Lambda_{0,0} - \sqrt{2}\Lambda_{0,\uparrow} - \Lambda_{+,+}) \\
         &\sqrt{2}\Lambda_{-,\uparrow} + 2\Lambda_{0,+} - \sqrt{2}\Lambda_{+,\uparrow} \\
         &\sqrt{2}\Lambda_{0,\downarrow} + -2\Lambda_{+,-} \\
         &\sqrt{2}\Lambda_{0,-} + 2\Lambda_{+,\downarrow} - \sqrt{6}\Lambda_{+,0} \\
         &\sqrt{2}\Lambda_{0,0} + \sqrt{6}\Lambda_{+,-} - \sqrt{6}\Lambda_{+,+} \\
         &\sqrt{2}\Lambda_{0,+} + \sqrt{6}\Lambda_{+,0} - 2\Lambda_{+,\uparrow} \\
         &\sqrt{2}\Lambda_{0,\uparrow} + 2\Lambda_{+,+} \\
     \end{aligned}$ &
     $\begin{aligned}                                               
         &-6\im\Lambda_{-,\downarrow} \\[.2em]
         &-4\im\Lambda_{-,-} \\[.2em]
         &-2\im\Lambda_{-,0} \\[.2em]
         &0 \\[.2em]
         &2\im\Lambda_{-,\uparrow} \\[.2em]
         &-4\im\Lambda_{0,\downarrow} \\[.2em]
         &-2\im\Lambda_{0,-} \\[.2em]
         &0 \\[.2em]
         &2\im\Lambda_{0,+} \\[.2em]
         &4\im\Lambda_{0,\uparrow} \\[.2em]
         &-2\im\Lambda_{+,\downarrow} \\[.2em]
         &0 \\[.2em]
         &2\im\Lambda_{+,0} \\[.2em]
         &4\im\Lambda_{+,+} \\[.2em]
         &6\im\Lambda_{+,\uparrow}
     \end{aligned}$ \\ \hline
    \end{tabular}}
    \caption{Action of $\mathcal{D}_a$ on $\Lambda_{m,\tilde{n}}$ for $j=0, ~1/2 ~\textrm{and}~ 1$.}
    \label{table4}
\end{table}

\end{document}